\documentclass[aps,pre,twocolumn,superscriptaddress,showkeys,nofootinbib,floatfix,amsmath,amssymb]{revtex4-2}

\usepackage{mathtools}
\usepackage{graphicx}
\usepackage{booktabs,threeparttable}
\usepackage{multirow}
\usepackage[dvipsnames]{xcolor}
\usepackage[breaklinks=true]{hyperref}
\makeatletter
\AtBeginDocument{%
  \def\p@subsection{}%         references: 3.2, not 3 3.2
  \def\p@subsubsection{}%
}
\makeatother

\let\parencite\cite

\newcommand{\be}{\begin{equation}}
\newcommand{\ee}{\end{equation}}
\newcommand{\bea}{\begin{align}}
\newcommand{\eea}{\end{align}}
\newcommand{\units}[1]{{~\rm #1}}
\newcommand{\fig}[1]{Fig.\,\ref{#1}}
\newcommand{\Fig}[1]{Figure\,\ref{#1}}

\newcommand{\figsand}[2]{Figs.\,\ref{#1} and \ref{#2}}
\newcommand{\eq}[1]{Eq.\,(\ref{#1})}
\newcommand{\Eq}[1]{Equation (\ref{#1})}

\newcommand{\eqsand}[2]{Eqs.\,(\ref{#1}) and (\ref{#2})}

\DeclareMathOperator{\sgn}{sgn}
\renewcommand{\Re}{{\rm Re}\,}

\begin{document}

\title{What Does FEXI Measure in Neurons?}

\author{Valerij G.\ Kiselev}
\email{kiselev@cbs.mpg.de}
\affiliation{Department of Neurophysics, Max Planck Institute for Human Cognitive and Brain Sciences,
  Stephanstrasse 1a, 04103 Leipzig, Germany}
\affiliation{Medical Physics, Department of Radiology, University Medical Center Freiburg,
  Faculty of Medicine, University of Freiburg, Killianstrasse 5a, 79106 Freiburg, Germany}
\author{Jing-Rebecca Li}
\affiliation{Inria-Saclay, Equipe Idefix, UMA, ENSTA Paris, Institut Polytechnique de Paris,
  91120 Palaiseau, France}

\date{November 14, 2026}

\begin{abstract}
Exchange between tissue compartments is crucial for the interpretation of
diffusion MRI measurements in brain gray matter. However, the reported
values of the exchange time are broadly dispersed, by about two orders of
magnitude. We analyze the measurement technique called Filter Exchange
Imaging (FEXI) using a numerical solution of the Bloch--Torrey equation in
digitized neurons downloaded from NeuroMorpho.org. The FEXI result, which
is the recovery of diffusion coefficient in cells with an impermeable
membrane, is multiexponential, with the time constants defined by the
eigenvalues of the Laplace operator. Fitting the commonly used exponential
recovery function results in a strong dependence of the apparent exchange
time $\tau_x$ on the mixing time interval involved as well as on the
adjustment or fixation of the equilibrium diffusivity. To obtain an
estimate of membrane permeability $\kappa$, we reinterpret previously
published data on the preexchange lifetime in neuronal cell culture. It
results in $\kappa\approx 0.004\units{\mu m/ms}$. The corresponding
exchange time $\tau_x\approx 140\units{ms}$. We hypothesize that previously reported -- an order of magnitude shorter --
exchange times  are due to fast geometric exchange inside the
ramified cells.
\end{abstract}


\maketitle

%%%%%%%%%%%%%%%%%%%

%------------------------------

\hspace{0.47\textwidth plus 0.5\textwidth minus 0.1\textwidth}
\noindent\raisebox{112mm}[0pt][0pt]{\color[rgb]{1,0,0}\parbox{0.6\textwidth}{This is a postprint of a paper accepted for publication in \href{https://direct.mit.edu/imag/article/doi/10.1162/IMAG.a.1500/140047/What-Does-FEXI-Measure-in-Neurons}{Imaging Neuroscience}. When citing it, please, refer to the \href{https://direct.mit.edu/imag/article/doi/10.1162/IMAG.a.1500/140047/What-Does-FEXI-Measure-in-Neurons}{journal version}.}}

%------------------------------

\section{Introduction}\label{sec_intro}

\begin{figure*}[tbp]
\includegraphics[width=0.99\textwidth]{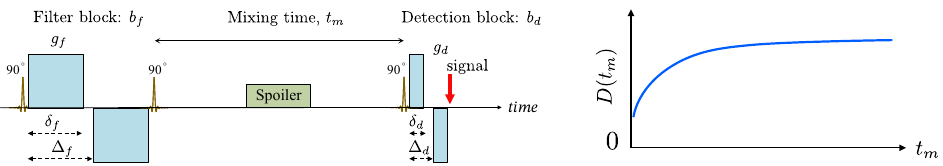}
\caption{Left: Schematics of FEXI as implemented in this study. Right: The anticipated recovery of diffusion coefficient with the increasing mixing time. 
}
\label{fig_sequence}
\end{figure*}

Diffusion MRI has been successfully applied to evaluate the microstructure of brain white matter~\cite{Novikov2019_models}. Extension of this approach to gray matter is challenged by exchange between intra- and extracellular space. Nulling the advantage of well-defined compartments, exchange becomes the primary focus of today's research. Among a number of approaches, filter exchange imaging (FEXI)~\cite{Aaslund2009,Lasic2011_exchange,Nilsson2013} appears a rather straightforward way to evaluate exchange between compartments with slow and fast diffusion, \fig{fig_sequence}.  However, this method  is sensitive to both the genuine exchange via the permeation through cellular membranes and to geometric exchange within the same cell~\cite{Qiao2005,Bernin2013,Nilsson2013,Khateri2022}. The latter takes place in cells with complex, tortuous geometry in which the filter, a strong diffusion weighting, does not affect nuclear spins with restricted mobility in the direction of the filter gradient. During the mixing time, these spins diffuse in regions inside the same cell without such restrictions thus contributing in the observed recovery of diffusion coefficient.  

In this paper, we scrutinize the outcome of FEXI in realistic neurons focusing on the geometric exchange inside the same cells. The cell membranes are first treated impermeable. We consider a simple model of long cylinder bent at the right angle to clarify the relation of the apparent exchange time to the eigenvalues of Laplace operator. On the next step, we find the membrane permeability by matching our in silico results for the preexchange lifetime with published experimental data~\cite{Yang2018}. The resulted value is used to simulate FEXI outcome in permeable cells, yet with idealized extracellular space. 

% Requires \usepackage{multirow} in the preamble (Table 2's multi-row
% header cells) and \usepackage{graphicx} (Table 1's \resizebox, used to
% force it to fit \textwidth). array is NOT used/needed anymore -- its
% m{} column type conflicts with revtex4-1's internal table macros.

% ============================================================
% TABLE 1: extended, wide version (5 cells), corrected data
% ============================================================
% Compression applied to make this fit \textwidth:
%  - \footnotesize content (smaller font -> narrower columns)
%  - \tabcolsep tightened to 0.25em (less padding between 16 columns)
%  - header parbox width trimmed 0.16\textwidth -> 0.115\textwidth
%  - whole tabular wrapped in \resizebox{\textwidth}{!}{...} as a safety
%    net: this guarantees an exact fit to \textwidth no matter what,
%    scaling the (already-compressed) table down further if still needed.
%    Requires \usepackage{graphicx} in the preamble.
\setlength\tabcolsep{0.25em}
\begin{table*}[tbp]
\footnotesize
\resizebox{\textwidth}{!}{%
\begin{tabular}{|c|lll|lll|lll|lll|}
\hline
 &\multicolumn{3}{|c|}{\href{https://neuromorpho.org/neuron_info.jsp?neuron_name=03b_spindle7aACC}{cell 1}} 
 & \multicolumn{3}{|c|}{\href{https://neuromorpho.org/neuron_info.jsp?neuron_name=02a_pyramidal2aFI}{cell 2}} 
 & \multicolumn{3}{|c|}{\href{https://neuromorpho.org/neuron_info.jsp?neuron_name=1-10-10}{cell 3}} 
 & \multicolumn{3}{|c|}{\href{https://neuromorpho.org/neuron_info.jsp?neuron_name=1-2-2}{cell 4}} \\
\hline
\parbox{0.2\textwidth}{\begin{flushleft} name, origin,\\ cell class, reference $\to$\end{flushleft} }&
\multicolumn{3}{|c|}{
\parbox{0.2\textwidth}{\begin{flushleft} 03b\_spindle7aACC,\\
neocortex, anterior cingulate, layer 5, von Economo neuron \parencite{Watson2006_neurons}\end{flushleft}}} &
\multicolumn{3}{|c|}{
\parbox{0.2\textwidth}{\begin{flushleft} 02a\_pyramidal2aFI,\\ neocortex, fronto-insula, layer 5, pyramidal neuron \parencite{Watson2006_neurons}\end{flushleft}}}
& \multicolumn{3}{|c|}{
\parbox{0.2\textwidth}{\begin{flushleft} 1-10-10,\\
neocortex, temporal, Brodmann area 22, pyramidal neuron \parencite{Jacobs2001_neurons}\end{flushleft}}}
& \multicolumn{3}{|c|}{
\parbox{0.2\textwidth}{\begin{flushleft} 1-2-2,\\
neocortex, temporal, Brodmann area 22, pyramidal neuron \parencite{Jacobs1997}\end{flushleft}}}\\
\hline
$V_{\rm cell}/V_{\rm soma}/f_{\rm soma}$~($\mu\mathrm{m}^3$/$\mu\mathrm{m}^3$) &
\multicolumn{3}{c|}{6397 / 5180 / 81\%} &
\multicolumn{3}{c|}{25629 / 20533 / 80\%} &
\multicolumn{3}{c|}{14407 / 11095 / 77\%} &
\multicolumn{3}{c|}{13573 / 9630 / 71\%} \\
\hline
$a$ &1 &2 &3    &1 &2 &3    &1 &2 &3    &1 &2 &3 \\
\hline
$e_x$ &0.0116 &0.6257 &-0.7800    &0.5890 &0.6728 &0.4477     &-0.2460 &0.9447 &0.2168     &-0.4111 &-0.5604 &0.7190 \\
$e_y$ &0.0134 &0.7799 &0.6258     &-0.0537 &-0.5202 &0.8523     &-0.1344 &-0.2547 &0.9576     &0.1004 &-0.8118 &-0.5752 \\
$e_z$ &0.9998 &-0.0177 &0.0006     &0.8064 &-0.5261 &-0.2703    &0.9599 &0.2064 &0.1897     &0.9061 &-0.1642 &0.3900 \\
\hline
$D^{(a)}/D_0$ &0.6632 &0.7585 &0.8378    &0.7621 &0.7990 &0.8687    &0.7106 &0.7321 &0.8291    &0.6738 &0.7387 &0.7654 \\
\hline
\end{tabular}%
}
\caption{Specification of human cells used in this study. Rows: cell description, the total and soma volumes, the number of eigenvector, $a=1,2,3$, eigenvectors of diffusion tensors (three columns per cell) and the corresponding normalized eigenvalues $D^{(a)}/D_0$.}
\label{tab_cells}
\end{table*}

\begin{figure*}[tbp]
\noindent
\includegraphics[width=0.990\textwidth]{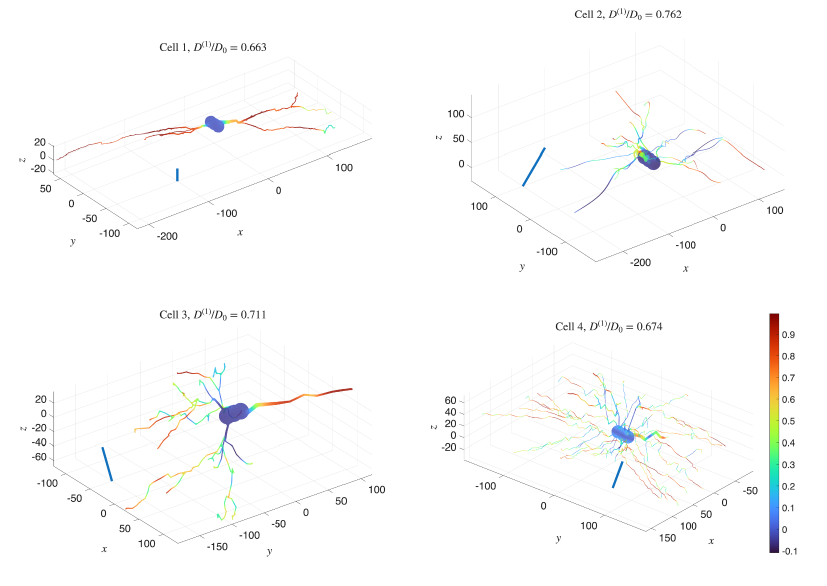} 
\caption{The four cells used in this study (Table\ \ref{tab_cells}). The colors encode the signal magnitude right after the filter. The filter gradient was applied along the eigenvectors with the smallest eigenvalues $D^{(1)}$, the directions shown with the blue lines. Videos of rotated cells are available via the \href{https://direct.mit.edu/imag/article/doi/10.1162/IMAG.a.1500/140047/What-Does-FEXI-Measure-in-Neurons}{journal version}. 
}
\label{fig_cells}
\end{figure*}

%%%%%%%%%%%%%%%%%%%%%
\section{Methods}\label{sec_methods}

\subsection{Digital Cells}

Four skeleton descriptions of human neurons were downloaded, in the SWC format, from the public repository NeuroMorpho.org~\cite{Tecuatl2024_NeuroMorpho,NeuroMorpho.Org}. Then {\verb|Alpha_Mesh_Swc|}~\cite{McSweeney-Davis2025}, an automatic and robust surface mesh generator for SWC files, was used to generate a three-dimensional surface mesh for each cell, which was then used by the tetrahedral mesh generator {\verb|tetgen|}~\cite{Si2015} to create a simulation-ready finite elements mesh.  
The total cell volumes and the soma volumes were calculated using the created surface mesh. Since the result disagree with the data from NeuroMorpho.org, we paid special attention to cross-checking the results, see Appendix \ref{ssec_cells}. The cells are shown in \fig{fig_cells}, and their properties are summarized in Table~\ref{tab_cells}.

\subsection{Numerical Solver}

The evolution of nuclear magnetization inside the cells was calculated with the diffusion MRI simulation software {\verb|SpinDoctor|}~\cite{Li2019,Fang2020}. It solves the Bloch--Torrey equation~\cite{Torrey1956} 
\be \label{BTeq}
\frac{\partial}{\partial t}\psi(t,{\bf r}) 
= D_0 \nabla^2 \psi(t,{\bf r}) - i{\bf g(t)}{\bf r}\, \psi(t,{\bf r}) \,
\ee
using the finite-element method.  Here the gradient of the Larmor frequency ${\bf g}$ and the position ${\bf r}$ are three-dimensional vectors and $\psi$ is the complex-valued transverse nuclear magnetization {\verb|SpinDoctor|} works with. To use it for FEXI, which involves all three magnetization components (\fig{fig_sequence}), we note that the $90^\circ$ rotation around the $y$-axis, spoiling the residual transverse magnetization and the back rotation to the transverse plane returns the original $x$-component of magnetization while nulling other component. In other words, it is taking the real part of the magnetization, $\psi \to \Re \psi$ after the filter block, which solves the problem. 

In this study, we do not take into account the longitudinal relaxation to focus on diffusion effects. When present, the longitudinal relaxation simply reduces $\Re \psi$ by an appropriate factor.

The boundary condition on the cell walls are routinely defined in terms of the membrane permeability $\kappa$ and the spin concentration $c$. The flux $-D{\bf n}\nabla c$ is continuous across the membrane, where ${\bf n}$ is the local normal vector of unit length. The concentration difference $\Delta c$ at the two membrane faces is defined from the condition $D{\bf n}\nabla c = \kappa \Delta c$. Note that the flux is zero for impermeable membranes with $\kappa=0$.   

The solution to \eq{BTeq} was found using the Matrix Formalism Method~\cite{Callaghan1997,Barzykin1999}, see also a detailed review~\cite{Grebenkov2007}. The core of the method is computing numerically a large number of eigenfunctions $\psi_n(x)$ of Laplace operator $D_0\nabla^2$ in the given cell geometry. After these functions and the corresponding eigenvalues $\lambda_n$ have been computed, the solution to \eq{BTeq} can be found for arbitrary diffusion-encoding sequence at little additional costs~\cite{Li2020}. To give a flavor of this solution, it implements the standard diffusion propagator for \eq{BTeq} when ${\bf g(t)}=0$. With an applied gradient, the operator ${\bf g(t)}{\bf r}$ is not diagonal in the eigenbasis, therefore it contributes to the propagator a matrix consisting of elements $\langle \psi_n \vert {\bf g(t)}{\bf r} \vert \psi_m \rangle$, where the bra-ket notation denotes the integral over the cell volume.

The numerical solution of the Bloch--Torrey equation was set to have $10^{-4}$ relative tolerance and $10^{-6}$ absolute tolerance for the ODE solver. The relative magnitude of the imaginary part of the signal, theoretically a real quantity, was about $10^{-5}$, in agreement with the predefined tolerance. The major numerical error was due to the limited size of finite elements mesh. For example, an about 10-fold increase in the mesh size in cell 2 (from $14\times 10^3$ to $142\times 10^3$) resulted in  a tolerable  change in $D(t_m)$ (the mixing time dependent diffusion coefficient defined below in \eq{Dapp_tm}), while significantly increasing the computation time. The change was 9\% for the shortest $t_m$ of $2\units{ms}$, monotonously decreasing to 0.5\% for $t_m=300\units{ms}$.

\subsection{Simulated FEXI}

The simulated FEXI weighting included the following parameters (\fig{fig_sequence}). Both the filter and the detection blocks used the Stejskal--Tanner diffusion weighting with rectangular gradient pulses of magnitude $g_f = g_d = 750\units{mT/m}$. For the filter, the gradient pulse duration was $\delta_f = 4\units{ms}$ with the interpulse interval $\Delta_f = 5\units{ms}$. This resulted in $b_f = 2.36\units{ms/\mu m^2}$. The timing of the detection block was $\delta_d = 1\units{ms}$ and $\Delta_d = 2\units{ms}$ ($b_d = 0.067\units{ms/\mu m^2}$).  Wide gradient pulses were used as the most realistic approach for the attempts to reach the maximum diffusion weighting in the shortest time. In the FEXI context,  this choice approaches the requirement of instant diffusion weighting, which largely simplifies the analysis. The bulk water diffusivity inside the cell was set to $D_0=2\units{\mu m^2/ms}$~\cite{Dhital2019}. The above parameters serve as the default unless stated otherwise. 

The direction of the filter and detection gradients were selected along the eigenvectors of cell's diffusion tensor $D_{ab}$. The tensors were found for each cell assuming impermeable membranes and using the detection block gradients applied in 32 directions distributed on a semi-sphere. It was important for the data consistency to use the same timing as in the detection block, since $D_{ab}$ sharply depends on time due to the contribution of narrow subcompartments within the cells. The eigenvectors and eigenvalues of $D_{ab}$ are summarized in Table~\ref{tab_cells}. The mixing time $t_m$ was the main variable in the simulations ranging from $5\units{ms}$ to $10\units{s}$ in the absence of any background relaxation.  
Such a long maximum time was selected to see the overall picture in the cells with the characteristic size of $100\units{\mu m}$. In other words, this time is commensurate with the reciprocal of the smallest nonzero eigenvalue of the Laplace operator. When presenting results, we pay special attention to the realistic values of $t_m$ up to few hundred of milliseconds.

To interpret the results,  we considered FEXI in a radically simplified geometry of a bent cylinder. It was created as a union of two equal cylinders joined at their ends at the right angle (\fig{fig_cells}). The radius of both parts was $\rho=1\units{\mu m}$ and the length $L/2$, with the total length $L=200\units{\mu m}$ roughly matching the maximum dimension of cells. 

\subsection{Evaluation of Membrane Permeability}

For the evaluation of cell membrane permeability, we used the SpinDoctor software to model the magnetization leakage from cells to extracellular space  where the signal was suppressed~\cite{Quirk2003,Bai2018,Yang2018}. The digitized cells were embedded in large rectangular boxes representing extracellular space. The initial magnetization there was zero and a very short $T_2=0.1\units{ms}$ simulated  the signal removal by mechanical perfusate flow  in the real experiment~\cite{Yang2018}. No inherent relaxation was assigned inside the cells. The initial magnetization there was spatially uniform. This approach was applied to reinterpret the most precise results of experiment in which the extracellular magnetization was mechanically removed by the perfusate flow~\cite{Yang2018}.

\subsection{FEXI in Permeable Cells}

The obtained membrane permeability was used to discuss the exchange with extracellular space (ECS) using a semi-analytical approach. The magnetization leakage in the large ECS was recalculated to find fractions of migrating spins for the realistic ECS volume using the Zimmerman--Brittin exchange model~\cite{Zimmerman1957}.  Each fraction was assigned a contribution in the overall diffusion coefficient according to its location during the application of the filter and detection blocks as explained below. 
 
%%%%%%%%%%%%%%%%%%%%%
\section{Results}
\subsection{Impermeable Cells}

\begin{figure*}[tbp]
\includegraphics[width=0.99\textwidth]{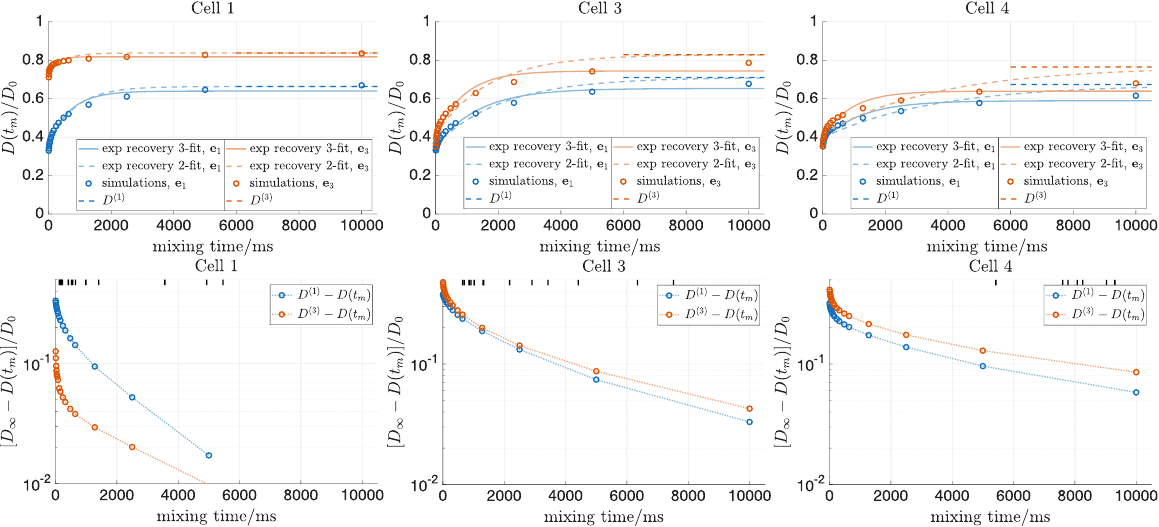}
\caption{The recovery of diffusion coefficient as a function of mixing time for three impermeable cells of increasing complexity. Diffusion-weighting gradients in the filter and detection block were applied along the eigenvectors, ${\bf e}_{1,3}$, with the smallest and largest eigenvalues, $D^{(1)}$ and $D^{(3)}$, respectively. 
Top row: Data obtained via the solution to the Bloch--Torrey equation (circles). The horizontal dashed lines indicate the undisturbed values of $D_\infty$, which are the corresponding eigenvalues of diffusion tensor. The dashed and solid lines in pale colors show $D(t_m)$, \eq{Dapp_tm}, fitted to data with all three parameters adjusted (``3-fit'' missing the asymptote) and with $D_\infty$ fixed at the correct values (``2-fit'' approaching the asymptote). Bottom row: The approach of $D(t_m)$ to its undisturbed values. The lines would appear straight if \eq{Dapp_tm} were correct. The ticks at the frame top show the time $\tau_n = 1/\lambda_n$ corresponding to 16 numerically found lowest nonzero eigenvalue of Laplace operator $-D_0\nabla^2$, \eq{cylinder_sol0}~\cite{Li2020}. The values $\tau_1=14\units{s}$ for cell 3 is outside the plot range. For cell4, $\tau_1=23\units{s}$ and the first one within the plot frame is $\tau_{10}=9.3\units{s}$. 
}
\label{fig_FEXI_cells}
\end{figure*}

The FEXI results for all four cells are shown in \fig{fig_FEXI_cells} along with the fitted recovery equation 
\be \label{Dapp_tm} 
D(t_m) = D_\infty - a\, e^{-t_m/\tau_x} \,,
\ee
 where $D_\infty$ is the equilibrium diffusivity, $a/D_\infty$ is the filter efficiency and $\tau_x$ the exchange time.  Here and in what follows, we denote exchange times with $\tau$ and the corresponding rates with $r=1/\tau$. The fitting was performed in two ways. For a two-parameter fit, $D_\infty$ was fixed at its correct values equal the corresponding eigenvalues of the diffusion tensor. For a three-parameter fit, all three parameter were adjustable.  The choice depends on a specific experiment. The two-parameter fit requires additional measurements with zero filter gradient. Such an option is excluded for the measurements in the strong static gradient produced by a single-sided permanent magnet~\cite{Williamson2019,Williamson2020,Williamson2025_preprint}. The  three-parameter fit is then inevitable.  A systematic deviation of the fitted lines from the data is evident for all cells and both ways of fitting. A pattern similar to the three-parameter fitting was reported in our previous simulations in porous media~\cite{Khateri2022}. 

\begin{figure*}[tbp]
\includegraphics[width=0.99\textwidth]{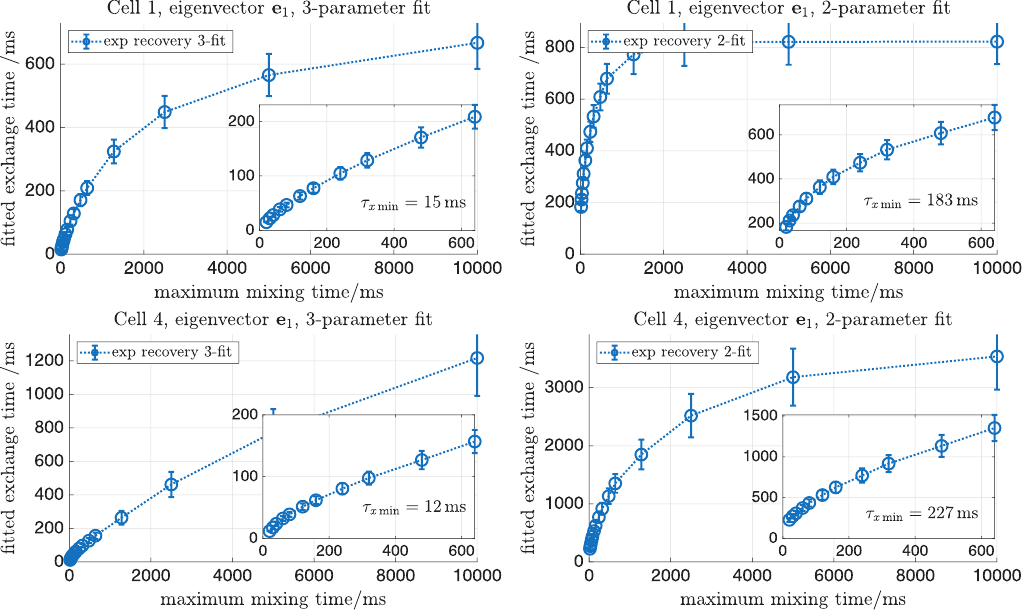}
\caption{Fitted apparent exchange time, $\tau_x$, as a function of the analyzed mixing time range for the simplest and the most developed cells. The fitting was performed from the minimum $t_m$ to the value indicated on the axis. The inset shows the practically relevant range of $t_m$. The error bars show the  fit error estimated from the fit residual. Left: Fitting all three parameters in \eq{Dapp_tm}. The first value for $t_m=20\units{ms}$ is $\tau_x=12\units{ms}$. Right: $D_\infty$ is fixed at its correct value. The first values for $t_m=20\units{ms}$ are given in the insets. 
}
\label{fig_cell_fit}
\end{figure*}

This deviation is visualized in \fig{fig_cell_fit} as the dependence of the apparent exchange time, $\tau_x$ on the interval of mixing times, $t_m$, used for fitting. For the two-parameter fit, fitting to the first three data points for $t_m = 5, 10, 20\units{ms}$ results in $\tau_x$ around $200\units{ms}$. This value increases to a few seconds for the whole interval of $t_m \leq 10\units{s}$. Of course, such long $t_m$'s are only available in simulations. Practically relevant are the values of a few hundred of milliseconds.  Over this interval, the obtained $\tau_x$ increases several-fold. 

The three-parameter fit delivers much shorter  exchange  times. Fitting to the first three data points results in  $\tau_x=12\mbox{--}15\units{ms}$. Using the full $t_m$ range up to $10\units{s}$ results in a few hundreds of milliseconds. Fitting within the practically relevant $t_m$ range results in an order-of-magnitude difference in $\tau_x$ between the short and long mixing time. 

\subsection{A Toy Model}

\begin{figure*}[tbp]
\includegraphics[width=0.99\textwidth]{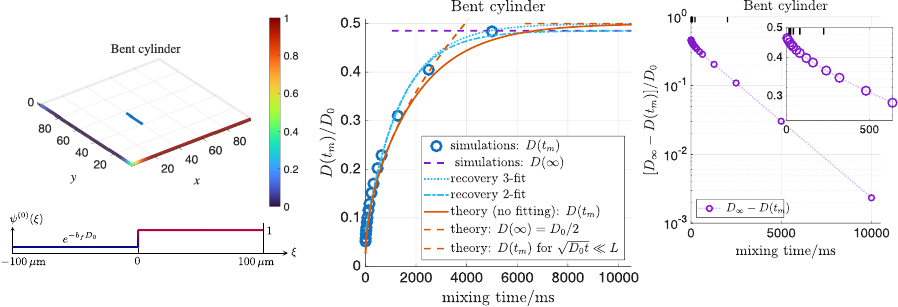}
\caption{The toy model of bent cylinder. Left: The cylinder drawn as in \fig{fig_cells} with the straightened presentation of the initial condition for the analytical solution, \eq{magnet_ini}.
Middle: Recovery of diffusion coefficient as in the top row of \fig{fig_FEXI_cells}. It is compared with the analytical result in the form of a series over the eigenvalues of the Laplace operator, \eq{Dapp} (red solid line). The theory can be simplified by assuming the approximation of \eq{magnet_ini}. The red dashed lines show the prediction of this simplified theory for the unperturbed diffusivity ($D_\infty=D_0/2$) and the short-time closed asymptotic form for $\sqrt{D_0 t}\ll L$, \eq{Dapp_short}. The dashed purple line shows the actual $D_\infty$, which is slightly smaller than $D_0/2$ due to the edge effect~\cite{Mitra1992}. The two-parameter fit (``2-fit'', dot-dashed light blue line) and three-parameter fits (``3-fit'', dotted light blue line) give close results, $\tau_x\approx 1600\units{ms}$ and $\tau_x\approx 1500\units{ms}$, respectively.
Right: The approach of $D(t_m)$ to its undisturbed asymptote as in the bottom row of \fig{fig_FEXI_cells}. The black ticks on the top have the same meaning as in \fig{fig_FEXI_cells} showing the first 6 time constants. 
}
\label{fig_FEXI_cylinder}
\end{figure*}

To interpret the above result, consider a simple model of a cylinder bent $90^\circ$ in the middle, \fig{fig_cells}. While this geometry is oversimplified as compared with neurons, it illustrates the essence of geometric exchange in an analytically tractable way. We first solved \eq{BTeq} numerically as for all cells. The result is  shown in \fig{fig_FEXI_cylinder}. Fitting \eq{Dapp_tm} demonstrates the same deviation pattern as for all cells. This similarity motivates scaling the analysis from the simplest model to cells, as discussed below.  

The simplest theory treats the filter and the detection blocks instant. The calculations are then focused on the magnetization evolution during the mixing time. This problem is effectively one-dimensional as long as $\rho\ll L$ and we use the coordiante $\xi$ for the whole cylinder length. Let the filter and the detection gradient be applied parallel to one of sections of the bent cylinder, say for $-L/2 < \xi <0$ (this section is $-y$ in \fig{fig_cells}). The evolution then starts with the  normalized  magnetizations 
\be \label{magnet_ini}
\psi_\parallel^{(0)} = e^{-b_f D_0}\,,\quad \psi_\perp^{(0)} = 1 \,
\ee
in the parallel and orthogonal sections, respectively. The approximation $\psi_\perp^{(0)} = 1$ indicates the neglect of diffusion across the thin cylinder. The first assignment is also an approximation that neglects the transition zone around $\xi=0$ and the edge enhancement, which can be barely recognized for $y\gtrsim 90\units{\mu m}$ in \fig{fig_cells}. The initial magnetization gradually evens out forming a smooth function $\psi(t_m,\xi)$. The detection block measures the diffusion coefficient $D_0$ in the parallel section and zero in the orthogonal one.  The overall diffusivity $D(t_m)$ is the weighted mean of these contributions with the weights  
\begin{align} 
S_\parallel(t_m) = \frac1L \int_{-L/2}^0 {\rm d} \xi\, \psi(t_m,\xi) \,, \label{Spara0}\\
S_\perp(t_m) = \frac1L \int_0^{L/2} {\rm d} \xi\, \psi(t_m,\xi) \,. \label{Sperp0}
\end{align}
In terms of these quantities, 
\be  
\frac{D(t_m)}{D_0} = \frac{S_\parallel(t_m)D_\parallel+S_\perp(t_m) D_\perp}{[S_\parallel(t_m)+S_\perp(t_m)]D_0} 
\approx \frac{S_\parallel(t_m)}{S_\parallel(t_m)+S_\perp(t_m)}  \,, \label{Dapp}
\ee
where we used the approximation $D_\perp=0$. 

The evolution of $\psi(t_m,x)$ during the mixing time is found routinely using the decomposition in the eigenfunctions of the Laplace operator, see Appendix \ref{ssec_sum}. The result is 
\begin{align} 
S_\parallel(t_m) &= \frac{\psi_\perp^{(0)}+\psi_\parallel^{(0)}}{4} \nonumber \\
&- \frac{4(\psi_\perp^{(0)}-\psi_\parallel^{(0)})}{L^2}  
\sum_{n=1}^\infty \frac{e^{-D_0k_n^2 t_m}}{k_n^2} \,, \label{Spara} \\
S_\perp(t_m) &= \frac{\psi_\perp^{(0)}+\psi_\parallel^{(0)}}{4} \nonumber \\
&+ \frac{4(\psi_\perp^{(0)}-\psi_\parallel^{(0)})}{L^2}  
\sum_{n=1}^\infty \frac{e^{-D_0k_n^2 t_m}}{k_n^2}  \label{Sperp} \,, 
\end{align}
where $k_n=(2n-1)\pi/L$. The exponential functions describe the leveling off of the initial magnetizations, $\psi_\perp^{(0)}$ and $\psi_\parallel^{(0)}$; the two expressions become equal at $t_m\to\infty$.  The terms with the series cancel out in the entire signal, $S(t_m)=(\psi_\perp^{(0)}+\psi_\parallel^{(0)})/2$,  for any time reflecting the absence of relaxation in the present consideration. 

The resulted diffusion coefficient, $D(t_m)$, which is calculated according to \eq{Dapp}, is shown in \fig{fig_FEXI_cylinder}. The agreement with the numerically calculated values is good giving the simplicity of the theory. For short times, $\sqrt{D_0 t_m}\ll L$,  the infinite series in \eqsand{Spara}{Sperp} can be effectively summed by using the free Gaussian propagator for the evolution of the magnetization formed right after the filter (appendix \ref{ssec_G}). This implied the neglect of the edge enhancement. The partial signals can then  be found in the closed form, \fig{fig_FEXI_cylinder} 
\be \label{Sboth}
S_{\parallel,\perp}(t_m) = \frac12 \psi_{\parallel,\perp}^{(0)} 
\pm (\psi_\perp^{(0)}-\psi_\parallel^{(0)})\sqrt{\frac{D_0 t_m}{\pi L^2}} \,.
\ee
\Eq{Dapp} then gives  
\be \label{Dapp_short}
\frac{D(t_m)}{D_0}  = \frac{\psi_\parallel^{(0)}}{\psi_\perp^{(0)}+\psi_\parallel^{(0)}} 
+ \frac{\psi_\perp^{(0)}-\psi_\parallel^{(0)}}{\psi_\perp^{(0)}+\psi_\parallel^{(0)}} \sqrt{\frac{4D_0 t_m}{\pi L^2}} \,.
\ee  
Note that this expression can accommodate a more precise form of magnetizations than the zero-order approximation given in \eq{magnet_ini}.  The result can be improved by accounting for diffusion in the orthogonal section. In particular, for short times the two terms in the numerator in the middle expression in \eq{Dapp} can be comparable, which promotes the effect of small diffusivity in the orthogonal section. Such an improvement is outside the scope of this paper as we focus on a qualitative understanding of diffusion recovery in neurons. 

The main lesson from the above analysis is that the recovery of diffusion coefficient  is substantially multiexponential.  It follows from the middle expression in \eq{Dapp} because the partial section signals, which are given in the form of series in \eqsand{Spara}{Sperp}, are the time-dependent weights for the averaging of two diffusion coefficients, $D_\perp$ and $D_\parallel$. It is also   evident in the bottom row in \fig{fig_FEXI_cells} and agrees with previous results~\cite{Ordinola2024}. This behavior contrasts with commonly implied \eq{Dapp_tm}, which describes the true transcytolemmal exchange provided a well-mixed magnetization inside the cell~\cite{Fieremans2010}. This discrepancy explains the poor fitting of \eq{Dapp_tm}. The interplay of many exponentials for short times can result in a specific functional dependence (e.g., $\sim \sqrt{t_m}$ for the bent cylinder, \eq{Dapp_short}). Note that this approximation might be sufficient for practical measurement in which $t_m$ is limited by the longitudinal relaxation. The crossover between the short and long times is set by the smallest nonzero eigenvalue of the Laplace operator, $t_m\sim 1/\lambda_1$, where $\lambda_1 = \pi^2D_0/L^2$ for the cylinder (it defines the first terms in the series in \eqsand{Spara}{Sperp}). For the considered parameters, $D_0=2\units{\mu m^2/ms}$ and $L=200\units{\mu m}$, this time is about $2000\units{ms}$, which is supported by data shown in \fig{fig_FEXI_cylinder} in which the short-time approximation ceases to work around this time. 

This consideration suggests the semilogarithmic scale for presenting the approach of diffusion coefficient to its asymptote, $D_\infty-D(t_m)$, which is shown in the bottom row of \fig{fig_FEXI_cells} for all cells. A similar dependence for the bent cylinder is shown in \fig{fig_FEXI_cylinder}. Initial rapid decrease in the difference $D_\infty-D(t_m)$ gradually tends to a straight line indicating the trend to the monoexponential attenuation, which is governed by the smallest nonzero eigenvalue of Laplace operator. This is very precise for the cylinder with its simple eigenvalue spectrum. The slope found from the data plotted in \fig{fig_FEXI_cylinder} is $-0.51\times 10^{-3}\units{ms^{-1}}$, which agrees well with the first nonzero eigenvalue $\lambda_1=\pi^2 D_0/L^2 = 0.49\times 10^{-3}\units{ms^{-1}}$. The above crossover time is the inverse of this quantity. Note that the next eigenvalue is 9 times larger, \eq{cyl_eigen}, corresponding to $220\units{ms}$ characteristic time, well in the range where the monoexponential attenuation sets up (\fig{fig_FEXI_cylinder}). This relatively early beginning of monoexponential relaxation explains the close results of two- and three-parameter fitting of \eq{Dapp_tm} for the bent cylinder, in contrast to the results for cells.

The monoexponential behavior for long times is not reached for cells within the first $10\units{s}$ due to more developed dendritic trees.  It can be explained by the presence of multiple small eigenvalues that result in large time constants, \fig{fig_FEXI_cells}. Pursuing this relation quantitatively is beyond the scope of this work because it requires the detailed geometry of eigenfunctions~\cite{Li2020}. 

\subsection{Membrane Permeability}

We focus now on experimental data for the preexchange lifetime in neural cells, which is straightforward to interpret. We denote this time with $\tau_{io}$, (the index implies the transition ``inside $\to$ outside''). The main idea of the measurement was to suppress the signal from extracellular space and monitor the decrease in the intracellular signal  due to the unidirectional magnetization leakage outside.  The suppression could be imposed by doping the extracellular space with a paramagnetic contrast agent~\cite{Quirk2003,Bai2018}. This method can be questioned due to possible interplay of the observed inversion recovery with the contrast agent-induced magnetic field inside the cells. This issue was absent for viable neural cell grown on glass microbeads for which the magnetization in extracellular space was removed with mechanical flow of perfusate~\cite{Yang2018}. The results of the three mentioned studies were $\tau_{io}=550\units{ms}$ in the rat brain~\cite{Quirk2003}, $r_{io} = 2.0\pm 1.7\units{s^{-1}}$ for the rat brain cortical cultures~\cite{Bai2018} and $\tau_{io}=750\pm 50\units{ms}$ for neurons and $\tau_{io}=570\pm 30\units{ms}$ for astrocytes~\cite{Yang2018}. With a reasonable agreement between each other, these values are more than an order of magnitude larger than the minimum exchange times obtained with diffusion MRI~\cite{Li2023,Chan2025}. 

For further analysis, we need to underscore the difference between the exchange time and preexchange lifetimes. Both notions rely on the standard exchange model between well-mixed compartments with  $f_i$ and $f_o$ being the amount of water in the intra- and extracellular compartment, respectively~\cite{Zimmerman1957},  
\be \label{eq_ex}
\begin{aligned} 
\frac{df_i}{dt} &= -r_{io}f_i + r_{oi}f_o  \\
\frac{df_o}{dt} &= +r_{io}f_i - r_{oi}f_o  \,.
\end{aligned}
\ee
Since the equilibrium occurs for equal concentrations, the transition rates obey the condition $v_i r_{io} = v_o r_{oi}$, where $v_i$ and $v_o$ are the volume fractions of two compartments. 

When used in the MRI context, these equation are supplemented with specific characteristics of two compartments such as  relaxation rates~\cite{Zimmerman1957} or diffusivities~\cite{Kaerger1985}. The quantities $f_i$ and $f_o$ become then the signal fractions, the fractions of water multiplied with its specific signal contribution.  For clarity, we keep the notation $f$ for the spin fractions and collect the gradient-related attenuation in additional factors. Experiments with the suppressed extracellular signal correspond to the case $f_o=0$, which means that the measured quantity is $r_{io}$. In contrast, diffusion MRI measures the bidirectional exchange with the rate $r_x = r_{io}+r_{oi}$~\cite{Callaghan2004} with the corresponding exchange time $\tau_x=1/r_x$. Using the equilibrium condition and the normalization, $v_i+v_o=1$,  
\be \label{excond}
r_x = r_{io}+r_{oi} = \frac{r_{io}}{v_o} \,.
\ee
The extracellular volume fraction $v_o$ lies in the range $0.20\mbox{--}0.25$ from electrophoretic measurements~\cite{Sykova2008}. The estimate of the intracellular volume fraction in the rat brain from the contrast agent suppression was $0.81\pm 0.08$~\cite{Quirk2003}. This value gives the smallest $\tau_x=v_o \tau_{io} $ with $v_o \approx 0.19$. The most precise result of direct measurements~\cite{Yang2018}, $\tau_{io} = 750\pm 50\units{ms}$ turns then into an estimate $\tau_x \approx 140 \units{ms}$. It remains significantly larger than the minimum exchange times, such as $\tau_x = 13\units{ms}$~\cite{Chan2025}. 

\begin{figure*}[tbp]
\includegraphics[width=0.99\textwidth]{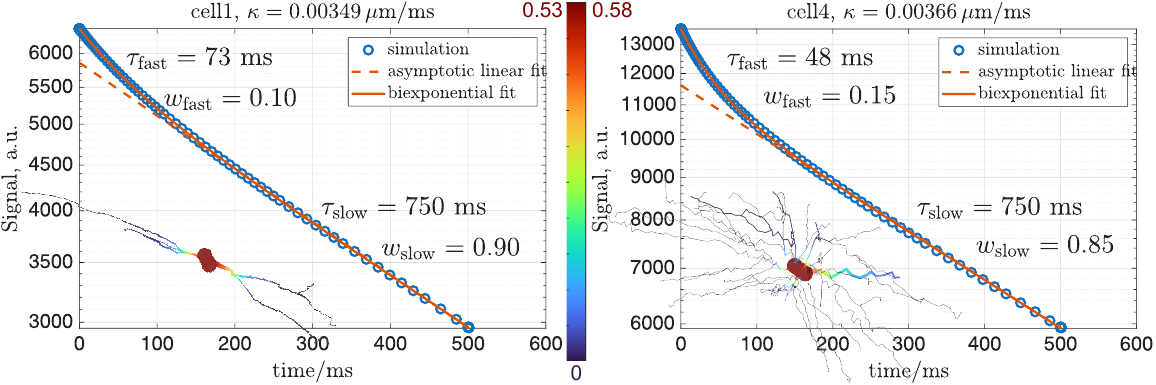}
\caption{Magnetization leakage from two cells to due to permeation in extracellular space in which $T_2$ is very short. Initial magnetization is unity, the signal is proportional to the cell volume. A biexponential function, \eq{biexp}, fitted to computed data yields the parameters indicated in the images. The insets show the cell's magnetization up to the end time with the same colormap as in \fig{fig_cells}. The colored numbers show the colormap range for both cells. The remaining magnetization in the soma is larger than in neurites. The membrane permeability was adjusted to match the measured preexchange lifetime~\cite{Yang2018}. 
}
\label{fig_escape}
\end{figure*}

To estimate the membrane permeability, we simulated the most precise preexchange lifetime measurement in neurons from the rat cerebral cortex~\cite{Yang2018}. \Fig{fig_escape} shows that the escape rate from the cells is not constant. In line with the original study, we fitted the data with a four-parametric biexponential function,  
\be \label{biexp}
S=S_0\cdot (w_{\rm fast} e^{-r_{\rm fast}t}+w_{\rm slow} e^{-r_{\rm slow}t})\,,
\ee
where the signal fractions are constrained to $w_{\rm fast} + w_{\rm slow} =1$. This is used to single out the long-time escape rate (the slow component). Beyond that, we do not speculate about the microstructural meaning of resulted parameters. Note that  a fast-escape component with $\tau_{\rm fast}\approx 50{\,\rm ms}$ was also observed in the original study. It was however attributed to extracellular space because it showed a dependence on the flow rate of perfusate. The slow component was independent, which suggested the interpretation as the genuine permeation. 

\begin{figure}[tbp]
\includegraphics[width=0.40\textwidth]{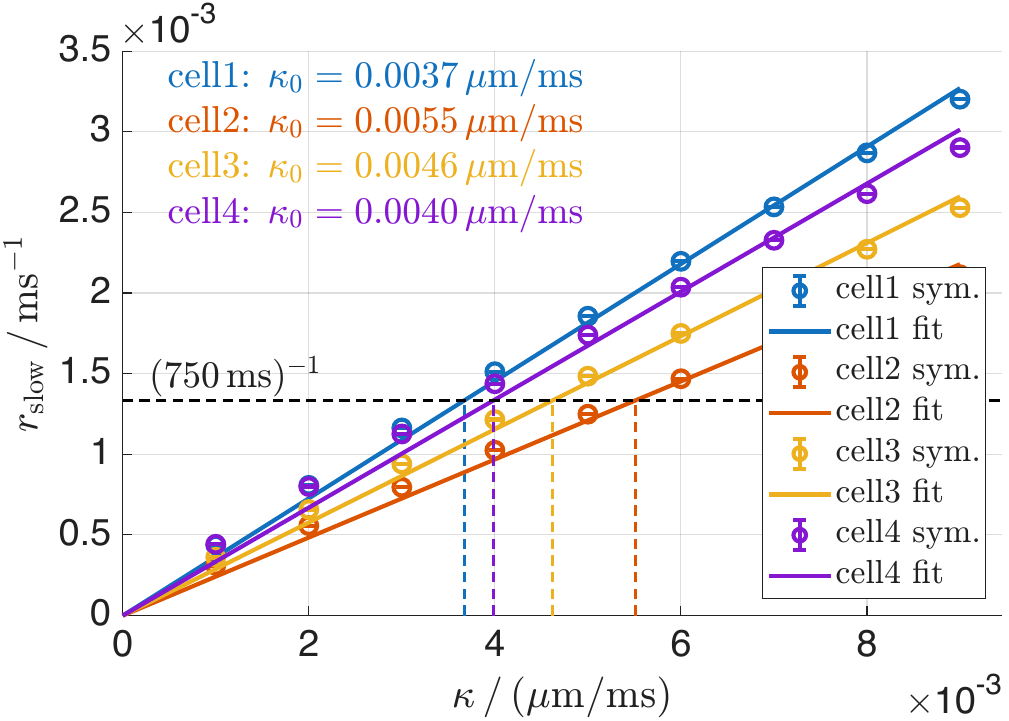}
\caption{The linear dependence between the long-time (slow) escape rate and the membrane permeability for four cells. The horizontal line shows the outcome of previous experiment~\cite{Yang2018}. Also shown are the values of membrane permeability that results in this value for all cells. The zero escape rate at $\kappa=0$ was forced in fitting. The present permeability for cells 1 and 4, are higher than the values indicated in \fig{fig_escape} because the fit to all nine values results in a line slightly below the accurate values near the crossing with the horizontal line. 
}
\label{fig_kappas}
\end{figure}

In line with that, we use the long-time exchange rate for interpreting our simulation results. As expected, this rate is proportional to the membrane permeability, \fig{fig_kappas}. Adjustment to the experimental value $r_{io} = (750{\,\rm ms})^{-1}$ $= 0.0013$ ${\,\rm ms^{-1}}$ requires the membrane permeability $\kappa = 0.0037,\,0.0055,\,0.0046,\,0.0040{\,\rm \mu m/ms}$ for the four cells, respectively \fig{fig_kappas}. 

We speculate that the admixture of extracellular space in the experiment~\cite{Yang2018} masked the true escape with $\tau_{\rm fast}$ close to our finding in the idealized in silico conditions. The plausible explanation of the fast component is a  rapid depletion of the magnetization in thin dendrites and in a layer adjacent to the soma membrane.  The following escape from the soma is slower being mediated by diffusion to the cell boundary and dendrites. This picture is supported by the higher magnetization in the soma up to the end of the considered time interval (\fig{fig_escape}). Note that the preexchange lifetime of around $50{\,\rm ms}$ agrees with the cylinder model having a typical dendrite caliber $\approx 1{\,\rm \mu m}$ and the membrane permeability $\kappa=0.004\units{\mu m/ms}$. 
However, the signal fractions of the fast component are smaller than the volume fractions of the neurites, cf.\ Table\ \ref{tab_cells}.  It prevents us from a too straightforward interpretation. 

Summarizing, we assume the membrane permeability $\kappa=0.004\units{\mu m/ms}$ in what follows.  It is a single-digit representation of the mean of four values shown in \fig{fig_kappas}. Such a low precision follows from the large uncertainty in the performed estimate. In particular, an  obvious drawback of our approach is the missing information about the cell geometry in the original measurements~\cite{Yang2018}.  It is a serious experimental challenge to perform experiments similar to those of Ref.~\cite{Yang2018}, but supplemented with quantifying the actual cell geometry.  

%%%%%%%%%%%%%%%%%%%%%
\subsection{FEXI in Permeable Cells}
\label{ssec_FEXIperm}

Simulating exchange in the whole neural tissue requires a number of cells occupying a volume fraction around 80\% with the rest volume attributed to extracellular space. Since we do not have such a comprehensive digitized tissue at hand, the addition of ECS occurs  to individual cells with some adjacent ECS. This is done in a semi-analytical way using  the exchange model of \eq{eq_ex} supplemented with the diffusion contributions of spins that depends on the spin locations during the application of the filter and the measurement blocks. In contrast to the K\"arger model~\cite{Kaerger1985}, diffusion-related signal attenuation is found numerically for spins experiencing either of these blocks inside the cell. This replaces the assignment of a specific diffusion coefficient in the K\"arger model. Similar to the K\"arger model, diffusion-related signal attenuation in the ECS is $e^{-bD_e}$ with the $b$-factor either $b_f$ or $b_d$ depending on the spin location during the application of the corresponding block. This approach neglects the spin migration during the block, which is a reasonable simplification for block duration much shorter than $t_m$. The extracellular volume fraction and the diffusivity were assumed to be $v_o=0.2$ and $D_e=0.8\units{\mu m^2/ms}$, respectively.  

The quantity of interest is now $f(t_m)=[f_i(t_m),$ $ f_o(t_m)]^\dagger$.  As discussed after \eq{eq_ex}, it consists of normalized spin fractions. The normalization implies $f_i(t_m)+f_o(t_m)=1$ for any $t_m$ in the absence of relaxation. In the presence of non-zero gradients, $f(t_m)$ acquires an explicit attenuation factor due to diffusion weighting. We denote such factors with $s$. 

For the analysis, we split $f_i(t_m)$ and $f_o(t_m)$ in  two spin populations each. This gives four groups of spins that contribute to the final diffusion coefficients $D(t_m)$. The groups are: (i) ``the residents'', which are initially in the cell and remain there until the end of the mixing time. Let the fraction of such spins in the overall spin pool be $f_{ii}(t_m)$ with the initial value $v_i$. Their contribution $D_i(t_m)$ to the final $D(t_m)$ is found numerically with account for the filter and the detection block; (ii) ``The emigrants'' that leave the cell to the ECS. Their spin fraction is $v_i-f_{ii}(t_m)$ with the filter-related attenuation, $s_i$, found numerically in the given cell geometry. The detection block yields the value $D_e$ because these spins are in the ECS by the moment of detection; (iii) ``The immigrants'' that move from the ECS into the cell. Since the overall spin density is constant, the fraction of this group is also $v_i-f_{ii}(t_m)$. The filter-induced attenuation is $s_e = e^{-b_f D_e}$. The detection block is applied to the intracellular magnetization $1-\psi(t_m,{\bf r})$, where $\psi(t_m,{\bf r})$ is the spin density (computed with no gradient) up to the end of the mixing time, $\psi(t_m,{\bf r})$ is shown in \fig{fig_escape}. This yields a numerical result for the diffusivity $D_i^{(0)}$; (iv) ``the foreigners'' that remain in the ECS. Their spin fraction is the original $v_o$ minus the ``immigrants'', $f_{ee}=v_o-f_{ei}$. The filter effect and the outcome of the detection block are both defined by $D_e$. The final $D(t_m)$ is the mean over the four groups weighted with their spin fractions and the filter effect, 
\be \label{Dtm}
D(t_m) = \frac{f_{ii} s_i D_i(t_m) + f_{ie} s_i D_e 
+ f_{ei} s_e D_i^{(0)} + f_{ee} s_e D_e}{f_{ii} s_i  + f_{ie} s_i  
+ f_{ei} s_e  + f_{ee} s_e} \, . 
\ee 
Note that $D_i^{(0)}$ is close to, but does not coincide with one of the eigenvalues $D^{(a)}$ when the gradient direction is aligned with the corresponding eigenvector of the diffusion tensor. The difference is due to the heterogeneous initial magnetization used in the calculation of $D_i^{(0)}$. 
Note that the described approach neglects exchange during the filter application and during the detection block, which is justified by their short duration. 

To find $f_{ii}(t_m)$,  we first use the analytical solution to \eq{eq_ex} with all spins initially localized inside the cell, $f(0)=[1,0]^\dagger$. It takes the formt
\be \label{exchange_M}
f(t_m) = \frac{1}{z+1}\begin{pmatrix}z+e^{-r_x t_m}\\ 1- e^{-r_x t_m}\end{pmatrix} \,,
\ee
where the top and bottom components are $f_i$ and $f_o$, respectively. Here $r_x$ is the exchange rate, \eq{excond}, and $z$ is the ratio of volume fractions, $z=v_i/v_o$. This solution takes into account exchange between intra- and extracellular space during the mixing time. 

Comparison of this solution with numerical calculations is not straightforward.  We use our simulation package to handle a cell embedded in a large box with fast relaxation, which implies $z\to 0$ in \eq{exchange_M}. In this limit, a spin that has left the cell has a vanishing return probability. \Eq{exchange_M} in this limit suggests that the exponential $e^{-r_x t}$ is the fraction of ``residents'', $f_{ii}$,  never leaving the cell, while the expression with finite $z$ takes  into account a finite return probability during the mixing time. 

Along this line, we find numerically the fraction of spins remaining in the cell embedded in a large box as in the setting yielding \fig{fig_escape}. So obtained quantity, $f_{ii}^{(0)}(t_m)$, replaces the exponential in \eq{exchange_M}. This gives the fraction of spins remaining in the cell with account for the reverse flow from a finite-volume, non-relaxing ECS,  
\be \label{Mi_tm}
f_{ii}(t_m)=\frac{z+f_{ii}^{(0)}(t_m)}{z+1}\,. 
\ee

\begin{figure*}[tbp]
\includegraphics[width=0.99\textwidth]{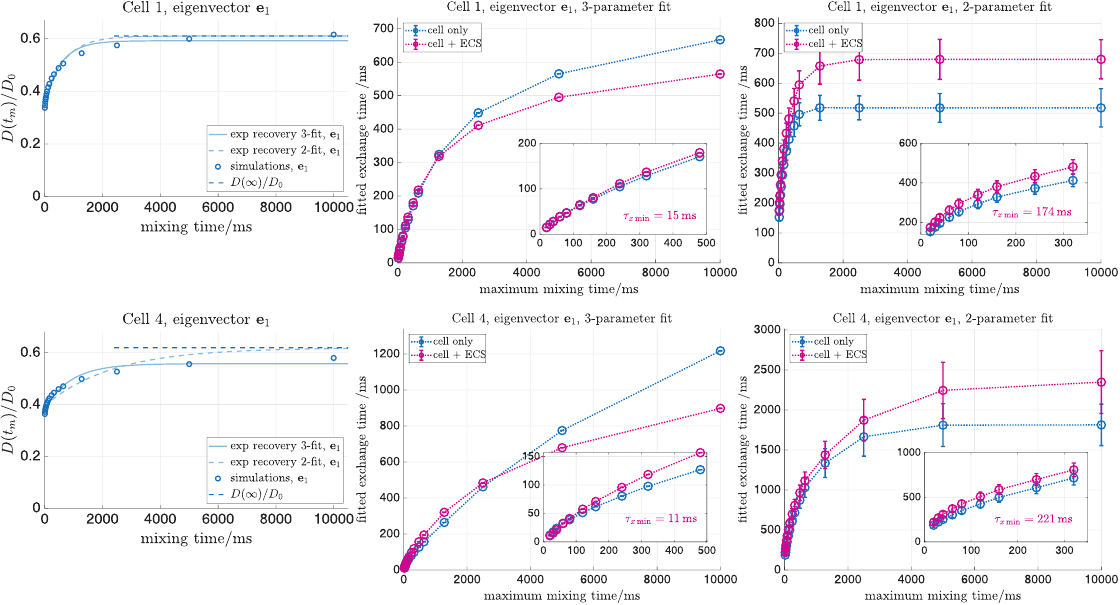}
\caption{Effect of exchange with added extracellular space on fitted apparent exchange time, $\tau_x$ for cell 1 (top) and cell 4 (bottom). The left column shows the recovery of diffusion coefficient as in \fig{fig_FEXI_cells}. Two middle and the right columns present the results for $\tau_x$ obtained with fitting \eq{Dapp_tm} to data with three and two adjustable parameters, respectively. The blue lines show the results for impermeable cells, and the purple lines show those for permeable cells. The minimum exchange time is shown for the permeable cells. The membrane permeability $\kappa=0.004\units{\mu m/ms}$. The error bars for the three-parameter fit are smaller than the symbol size. 
}
\label{fig_FEXI_perm}
\end{figure*}

The resulted diffusion coefficient,  \eq{Dtm}, is subjected to the multiexponential recovery similar to the behavior for impermeable cells (\fig{fig_FEXI_perm}). The dependence of the apparent exchange time on the interval of fitted data shows different patterns. The tree-parameter fit of \eq{Dapp_tm} is flexible enough to provide for a high precision although missing the correct long-time asymptote. In contrast, \eq{Dapp_tm} with $D_\infty$ fixed at its correct value is too rigid to fit the simulated data. The reason is that the approach of data points to the asymptote is very slow, which is a manifestation of a number of large time constants $1/\lambda_n$ (there are nine such constants above $10\units{s}$ in cell 4 vs.\ one only in cell 1).

Consider now the relation between the apparent exchange time in impermeable vs.\ permeable cells. Common sense suggests that an additional exchange channel should shorten the exchange time. It is so for cell 1 with the three-parameter fit. However, we cannot trust a fit function, \eq{Dapp_tm}, whose functional form does not match the physics. The common sense expectation is violated for cell 4 and -- in a very pronounced way --  for both cells when the two-parameter fit is used. The origin of this behavior is evident from the bottom-left panel in \fig{fig_FEXI_perm}. Forcing the fitted curve to the asymptote, $D_\infty$, is compensated with a reduction for short $t_m$'s. This mimics a larger $\tau_x$ in \eq{Dapp_tm} in comparison with the three-parameter fit.

%%%%%%%%%%%%%%%%%%%%%
\section{Discussion}

\subsection{Mismatch in Exchange Times}

Exchange time $\tau_x$ has been targeted in a number of diffusion MRI experiments. The results are widely, by two orders of magnitude, dispersed, see Fig.\,1 and Table\,S1 in Ref.~\cite{Li2023} for an overview. An obvious reason for such a dispersion is the variety of objects, methods and targeted quantities. Focusing on the fast-exchange side of the reported spectrum, the exchange time reaches a few tens of milliseconds. In particular, a thorough recent study~\cite{Chan2025} reported the median value $\tau_x=r_x^{-1}=13{\,\rm ms}$ (using an other method, not FEXI). This time is surprisingly short, which motivates a deeper look into different measurement techniques.  In what follows, we discuss implications of our results for real neuronal cells, while keeping in mind a number of limitations of our study. 

Two central quantity governing exchange are the membrane permeability and the surface-to-volume ratio of cells. Our discussion focuses on the fact that the obtained estimate of the permeability is rather low, $\kappa = 0.004\units{\mu m/ms}$, about an order of magnitude smaller than the erythrocyte value cited below. Further, the exchange time $\tau_x=140\units{ms}$ is an order of magnitude longer than the reported experimental results~\cite{Li2023,Chan2025}. 

\subsection{Membrane permeability}

Direct measurements of membrane permeability are scarce. To get an idea about the actual values, erythrocytes, which supposed to be the most permeable cells, have a membrane with $\kappa = 0.06\mbox{ --- }0.07\units{\mu m /ms}$ at the body temperature~\cite{Benga1988,Benga1990}. Neural cells were subjected to the osmotic challenge in a few studies. The results are controversial, from undetectable~\cite{Andrew2006} to the value $\kappa = 0.047{\,\rm \mu m/ms}$~\cite{Boss2013}. 

%%%%%%%%%%%%%%%%%%%%%

\subsection{Exchange time}

How to reconcile the broad band of experimental finding for the exchange time? The experimental setup with the suppression of extracellular signal reflects the {\it genuine membrane permeation}. In contrast, diffusion-based measurements such as FEXI are  sensitive to any change in the local diffusivity of spins, in other words to both the true permeation and to geometric exchange. The relative contribution of these processes depends on the mixing time sampled in an experiment. For short times, the fastest process dominates. According to our interpretation of cell culture experiment~\cite{Yang2018}, the true permeation has a characteristic time about $140\units{ms}$. Therefore, an order of magnitude shorter exchange times, such as $\tau_x \approx 13\units{ms}$~\cite{Chan2025} should be attributed to the {\it geometric exchange inside} ramified cells. 

Of course, this is currently a hypothesis, since the short exchange time was obtained with NEXI or SMEX~\cite{Jelescu2022_NEXI,Olesen2022}, the standard model of white matter~\cite{Novikov2019_models} supplemented with exchange. Hypothesis validation for this method is the topic of nearest future.

Meanwhile, we cannot exclude a possibility that the actual surface-to-volume ratio of real neurons is substantially larger than for the digitized models used in this work. In such a case, a low membrane permeability can be compensated by a large exchange surface. This increase can be mediated by spines~\cite{Chakwizira2026}, which are absent in the considered digitized cells. Since spines are small, diffusion coarse-grains such feature during the characteristic time of a few milliseconds. However, very thin spine necks are able to slow down the dendrite-spine exchange~\cite{Chakwizira2026,Simsek2026} thus shifting the effective coarse graining to later times. The coarse-graining results in renormalization of effective dendrite size, diffusivity and the membrane permeability. Due to technical limitations, investigating this effect remains a topic for future work.

As discussed in the context of \fig{fig_escape}, the thin dendrites mediate fast exchange. The relation between the soma and dendrite volumes is thus a relevant parameter for the exchange rate. The cells considered here have a relatively large soma, which takes $81\%$, $80\%$, $77\%$ and $71\%$ of volume for the four cells, respectively (Table~\ref{tab_cells}). 
This might be due to practical limitation in quantifying the cell shape. We cannot also exclude that cells grown on microbeads in the previous experiments~\cite{Yang2018} had changed their morphology in comparison with their in vivo state. 

The signal fractions in soma and dendrites is obviously affected by the strength of diffusion weighting. The filter strength considered here ($b_f = 2.36\units{ms/\mu m^2}$) is moderate in comparison with the performance of advanced scanners. Stronger weighting will promote the dendritic contribution to the signal with its faster exchange~\cite{Olesen2022,Jelescu2022_NEXI,Chan2025}.

Another approach to the exchange evaluation is the observation of the time-dependent kurtosis $K(t)$. In idealized systems with multiple permeable restrictions, kurtosis is zero for very short time because of originally Gaussian diffusion and for very long times due to the coarse graining~\cite{Novikov2010}. There is a peak in between reflecting the non-Gaussian nature of diffusion for intermediate times. The peak time $t_{\rm peak}$ can be found in the adiabatic approach, when diffusion inside individual compartments provides for a much faster mixing than the permeation between compartments. For a dense random packing of spheres with radius $\rho$~\cite{Lee2025}, 
\be \label{tpeak}
t_{\rm peak} = \left(\frac{2\rho^3}{5\kappa D_0} \right)^{1/2} \,,
\ee
where $D_0$ refers to the diffusivity outside the spheres. 

Straightforward application of this result speaks against our estimate of $\kappa=0.004\units{\mu m/ms}$. Indeed, for this value, $D_0 = \units{3 \mu m^2/ms}$ (free diffusion) and a relatively small effective cell radius $\rho = 3 \units{\mu m}$, $t_{\rm peak}$ already exceeds 25 ms. This value would be even larger for bigger cells and would further increase for smaller $D_0$. However, most measurements of time dependent mean kurtosis in gray matter suggest shorter $t_{\rm peak}$'s. The reported upper bounds are $10\units{ms}$~\cite{Jelescu2022_NEXI}, $12\units{ms}$~\cite{Li2023} in rodents,  $20\units{ms}$~\cite{Lee2020_GM} and $12.5\units{ms}$~\cite{Hao2025} in humans. Note that a hint of a maximum around $20\units{ms}$ is seen in the latter study, but only in the thalamus. 

We argue that the above theory, \eq{tpeak}, does not take into account the ramification of cells. Hypothetically, the dendrites can serve as exchange mediators due to their small diameter. Indeed, the dendrite radius of $1\units{\mu m}$ (somehow extended to account for spines), result in $t_{\rm peak} = 12\units{ms}$ according to \eq{tpeak}. This is supported by the large difference between the cell-escaping magnetization in soma and dendrites, which is visible in \fig{fig_escape}. Further reasons for the role of dendrites is their appreciable volume fraction, about $1/3$~\cite[and references therein]{Karbowski2015}, which is larger than in the cells used in this study, and the magnetization transport from the soma towards dendrites. 

One cannot exclude a simpler reason for not observing the $K(t)$ peak. Difference in intra- and extracellular diffusivity increases kurtosis at small times, which might mask the peak. Also contributing to the uncertainty might be the intravoxel variability in tissue parameters, which is partially taken into account by the introduction of $K_\infty$ for long times~\cite{Lee2020_GM}. Summarizing the above discussion, the membrane permeability remains yet elusive thus challenging future experimental and theoretical work. 

\subsection{On clinical feasibility}
The measurement parameters used in this study are typical for animal scanners. The major challenge for translation to clinical scanners is the limited gradient strength available for the filter block. Consider, for example, an advanced clinical scanner with a maximum gradient strength of $80\units{mT/m}$. Achieving a filter with $e^{-b_f D_0}=0.01$ and a $1\units{ms}$ gap for the refocusing pulse requires a long weighting time of $\Delta_f = 20\units{ms}$. The filter efficiency is reduced by geometric exchange that already takes place during this time. As an example, $D(t_m)$ increases by 23\% between $t_m = 20\units{ms}$ and $480\units{ms}$ within cell 4 along the $D^{(1)}$ direction (\fig{fig_FEXI_cells}). The long filter duration reduces this effect to a factor of 1.12 (data not shown). A weaker filter with $e^{-b_f D_0}=0.1$ requires $\Delta_f = 16\units{ms}$, and the increase in $D(t_m)$ is then a factor of 1.07. In this case, a stronger filter is preferable to a shorter one.  

Evaluating the measurement feasibility on clinical scanners requires an accurate estimate of the effect-to-noise ratio that accounts for transverse and longitudinal relaxations. Such an analysis is not performed here because it would go far beyond the scope of the present study.

\subsection{Outlook}

Focusing on FEXI, a way to validate the present results could be an observation of essentially multiexponential recovery of the filter-suppressed diffusion coefficient, similar to the simulation results shown in \figsand{fig_FEXI_cells}{fig_cell_fit}. 

Another manifestation of the multiexponential recovery of diffusion coefficient in FEXI is a strong dependence of the fitted exchange time on the number of free parameters in \eq{Dapp_tm}. This effect can contribute to the dispersion of reported $\tau_x$ values.

%%%%%%%%%%%%%%%%%% Ending sections (unnumbered)

\subsection*{Data and Code Availability}  
As stated above, the digitized cells are available from \href{https://neuromorpho.org}{NeuroMorpho.org}. The hyperlinks to the individual cells are given behind the title row in Table~\ref{tab_cells}. The SpinDoctor package is publicly available on \href{https://github.com/SpinDoctorMRI/SpinDoctor}{https://github.com/SpinDoctorMRI/SpinDoctor}. The control files for the specific tasks of this study are available on request, provided the requester acknowledges this in the resulting publication. The reason is that the simulation and presentation files are deeply intertwined with other functions on local computers. 

\subsection*{Funding}
This work was supported by the Engineering for Health (E4H) interdisciplinary center of Institut Polytechnique de Paris, and the ``Biomedical Engineering Seed Grant Program'' funded by the Fondation Bettencourt Schueller.  

\begin{acknowledgments}
We are grateful to our colleagues for stimulating discussions: Joseph J.\ Ackerman, Sune Jespersen, Markus Nilsson, Dmitry S.\ Novikov, and Nathan H.\ Williamson.
\end{acknowledgments}

%%%%%%%%%%%%%%%%%%
\appendix

\section{Cell properties}
\label{ssec_cells}

We calculated the total cell volume based on the created surface mesh using MATLAB and Python. The tetrahedral finite-element meshes exactly matched the surface facets, while the inner mesh points were independent. The surface matching explains the exact agreement between the two computations. 

In contrast, the results for the soma volumes agree, but not precisely. The reason for the deviations is the ambiguity in defining the boundaries between the soma and the cell processes. 

An additional independent calculation of soma metrics was based on the observation that all somas are approximated with three spheres. The surface and the volume of overlapping spheres were calculated analytically and implemented in MATLAB. It shows a good agreement with the other methods, but not with the NeuroMorpho data. Summarizing, the results based on finite-element mesh processed in MATLAB were used in this study as presented in Table \ref{tab_cells}. 

% Requires \usepackage{multirow} in the preamble and \usepackage{graphicx}
% (for \resizebox, used to force the table to fit \textwidth).
% ============================================================
% Data provenance (see table notes below for details):
%   NeuroMorpho        -- statistics from the NeuroMorpho.org entry for
%                          each reconstruction (automated L-Measure-type
%                          statistics computed from the raw SWC skeleton).
%   Mesh (FEM), MATLAB -- volumes computed from the tetrahedral finite
%                          element mesh used in the diffusion simulations,
%                          via a live MATLAB run of
%                          compute_cell_soma_volumes.m (ecs_shape =
%                          "no_ecs"; soma separated via separate_soma.m).
%   Mesh (FEM), Python  -- the same mesh/segmentation pipeline,
%                          independently reproduced in Python (TetGen +
%                          NumPy), used to cross-check the MATLAB result
%                          and to obtain soma surface area (which
%                          compute_cell_soma_volumes.m does not compute).
%   3-sphere model      -- analytic model of the soma as three pairwise-
%                          overlapping spheres, centered at the three
%                          SWC soma nodes, from a live MATLAB run of
%                          find_soma_spheres.m.
\setlength\tabcolsep{0.35em}
\begin{table*}[tbp]
%\footnotesize
\begin{tabular}{|l|l|r|r|r|r|}
\hline
Quantity & Method & cell 1 & cell 2 & cell 3 & cell 4 \\
\hline
\multirow{3}{*}{Total cell volume ($\mu\mathrm{m}^3$)}
  & NeuroMorpho              & 1605  & 7556  & 3556  & 4668  \\
  & Mesh (FEM), MATLAB$^{a}$ & 6397  & 25629 & 14407 & 13573 \\
  & Mesh (FEM), Python$^{b}$ & 6397  & 25629 & 14407 & 13573 \\
\hline
\multirow{3}{*}{Soma volume ($\mu\mathrm{m}^3$)}
  & Mesh (FEM), MATLAB$^{a}$ & 5180  & 20533 & 11095 & 9630  \\
  & Mesh (FEM), Python$^{b}$ & 5184  & 20572 & 11090 & 9633  \\
  & 3-sphere model$^{c}$     & 5172  & 19944 & 10958 & 9571  \\
\hline
\multirow{3}{*}{Soma surface ($\mu\mathrm{m}^2$)}
  & NeuroMorpho              & 812   & 1997  & 1340  & 1224  \\
  & Mesh (FEM), Python$^{b}$ & 1579  & 4060  & 2726  & 2533  \\
  & 3-sphere model$^{c}$     & 1625  & 3996  & 2680  & 2449  \\
\hline
\end{tabular} 
\caption{Cell metrics computed as described in the text. Note the difference between the NeuroMorpho data and the results calculated here. FEM = the tetrahedral finite-element mesh. }
\label{tab_cells_stat}
\end{table*}

\section{Magnetization during the mixing time}
\label{ssec_sum}

We describe here the evolution of longitudinal magnetization during the mixing time. The problem is effectively one-dimensional as $\rho\ll L$. The general solution is 
\be \label{cylinder_sol0} 
\psi(t_m,\xi) = \sum_{n=0}^\infty C_n \psi_n(\xi) e^{-\lambda_n t} \,,
\ee
where $\psi_n(x)$ are the eigenfunctions of the Laplace operator $-D\,\partial^2/\partial x^2$ with the eigenvalues $\lambda_n$ and the coefficients $C_n$ being the projections of the initial magnetizations on $\psi_n(\xi)$. For the problem at hand, $\xi$ is the straitened distance along the cylinders, $-L/2<\xi<L/2$ (\fig{fig_FEXI_cylinder}).  

The eigenfunctions are well known, 
\begin{align}
\psi_0(\xi)  &= \sqrt{{1\over L}}\,,\  \lambda_0=0 \nonumber \\
\psi_n(\xi) &= \sqrt{{2\over L}} \sin k_n \xi \,,\quad  \nonumber 
k=\frac{(2n-1)\pi}{L}\,,\  \lambda_n=D_0 k_n^2 \,, \\
n&=1,2,3\dots    \label{cyl_eigen} 
\end{align}
Not written here are the even functions, proportional to $\cos p_n \xi$. These function do not enter the result by symmetry and the orthogonality to $\psi_0$.   The deviation of the initial condition from an odd function is fully accounted for by the contribution of $\psi_0(\xi)$. 

The initial conditions read to the first approximation 
\be \label{magn_ini_all}
\psi(0, x) = \left\{
\begin{array}{ll}
e^{-b_f D_0} & 
\mbox{ for } -{L\over 2} <\xi<0 \\
1 &\mbox{ for } 0<\xi<{L\over 2} 
\end{array} \,. \right.
\ee
according to \eq{magnet_ini}.
To use the symmetry of the bent cylinder, we express this initial magnetization in terms of the even and odd components,  
\bea 
\psi_{\rm even}^{(0)} &= \frac{1+e^{-b_f D_0}}{2} \\
\psi_{\rm odd}^{(0)} &= \frac{1-e^{-b_f D_0}}{2} \sgn \xi \,,
\end{align}
where $\sgn \xi = \xi/|\xi|$ is the sign of $\xi$. The initial condition is then $\psi(0,\xi) = \psi_{\rm even}^{(0)}+\psi_{\rm odd}^{(0)}$. By the linearity of diffusion equation, the solution $\psi(t_m,\xi)$ is the sum of two solutions with the initial conditions $\psi_{\rm even}^{(0)}$ and $\psi_{\rm odd}^{(0)}$. The former excites the constant $\psi_0$ only. The odd part excites all odd modes, \eq{cyl_eigen}, with $C_n=\sqrt{2/L}/k_n$. Summarizing, 
\be \label{psi_cylinders} 
\psi(t_m,\xi) = \psi_{\rm even}^{(0)} 
+ |\psi_{\rm odd}^{(0)}| \frac{4}{L}\sum_{n=1}^\infty \frac{\sin k_n \xi}{k_n} e^{-D_0 k_n^2 t_m} \,.  
\ee
Separate integrations over the two section result in \eqsand{Spara}{Sperp}.

\section{Short-time approximation}
\label{ssec_G}

The general solution, \eq{cylinder_sol0}, can be written in terms of the propagator (the Green's function) of diffusion equation, $G(t_m,x,x_0)$,  
\be \label{cylinder_solG}
\psi(t_m,\xi) = \int_{-L/2}^{L/2} {\rm d}\xi_0 \, G(t_m,\xi,\xi_0) \psi(0,\xi_0) \,.
\ee
This exact solution becomes approximate when the propagator is treated as the free one 
\be 
G(t_m,\xi,\xi_0) \approx G_0(t_m,\xi-\xi_0) = \frac{e^{-(\xi-\xi_0)^2/(4D_0 t_m)}}{\sqrt{4\pi D_0 t_m}} \,,
\ee
which is possible for $\sqrt{D_0 t_m}\ll L$.  The substitution of $\psi(t_m,\xi)$ in \eqsand{Spara0}{Sperp0} results in two-dimensional integrals that are calculated in the limit $L\to\infty$. This  results in \eq{Sboth}.

%%%%%%%%%%%%%%%%%%
\onecolumngrid   % balances the two columns of the current page
\clearpage       % place all pending floats 
\twocolumngrid   % references in two columns
\bibliographystyle{apsrev4-2}
\bibliography{literature}

@article{Aaslund2009,
	author = {Aslund, Ingrid and Nowacka, Agnieszka and Nilsson, Markus and Topgaard, Daniel},
	doi = {10.1016/j.jmr.2009.07.015},
	journal = {J Magn Reson},
	journal-full = {Journal of magnetic resonance (San Diego, Calif. : 1997)},
	mesh = {Algorithms; Cell Membrane; Cell Membrane Permeability; Diffusion; Magnetic Resonance Spectroscopy; Saccharomyces cerevisiae; Water},
	month = oct,
	number = {2},
	pages = {291-5},
	pmid = {19647458},
	pst = {ppublish},
	title = {Filter-exchange {PGSE NMR} determination of cell membrane permeability},
	volume = {200},
	year = {2009}}

@article{Andrew2006,
	author = {Andrew, R. David and Labron, Mark W. and Boehnke, Susan E. and Carnduff, Lisa and Kirov, Sergei A.},
	doi = {10.1093/cercor/bhk032},
	journal = {Cerebral Cortex},
	month = {05},
	number = {4},
	pages = {787-802},
	title = {Physiological Evidence That Pyramidal Neurons Lack Functional Water Channels},
	volume = {17},
	year = {2006}}

@article{Bai2018,
	author = {Bai, Ruiliang and Springer, Jr, Charles S and Plenz, Dietmar and Basser, Peter J},
	doi = {10.1002/mrm.26980},
	journal = {Magn Reson Med},
	journal-full = {Magnetic resonance in medicine},
	mesh = {Animals; Biological Transport, Active; Body Water; Cell Membrane; Cells, Cultured; Cerebral Cortex; Models, Biological; Neurons; Ouabain; Rats; Rats, Sprague-Dawley; Sodium-Potassium-Exchanging ATPase},
	month = {06},
	number = {6},
	pages = {3207--3217},
	pmid = {29106751},
	pst = {ppublish},
	title = {Fast, {Na+ /K+} pump driven, steady-state transcytolemmal water exchange in neuronal tissue: {A} study of rat brain cortical cultures},
	volume = {79},
	year = {2018}}

@article{Barzykin1999,
	author = {Barzykin, A.V.},
	doi = {10.1006/jmre.1999.1778},
	issn = {1090-7807},
	journal = {Journal of Magnetic Resonance},
	number = {2},
	pages = {342-353},
	title = {Theory of Spin Echo in Restricted Geometries under a Step-wise Gradient Pulse Sequence},
	url = {https://www.sciencedirect.com/science/article/pii/S1090780799917780},
	volume = {139},
	year = {1999}}

@article{Benga1988,
	author = {Gheorghe Benga},
	doi = {10.1016/0079-6107(88)90002-8},
	issn = {0079-6107},
	journal = {Progress in Biophysics and Molecular Biology},
	number = {3},
	pages = {193--245},
	title = {Water transport in red blood cell membranes},
	url = {https://www.sciencedirect.com/science/article/pii/0079610788900028},
	volume = {51},
	year = {1988}}

@article{Benga1990,
	author = {Benga, G and Pop, V I and Popescu, O and Borza, V},
	doi = {10.1016/0165-022x(90)90057-j},
	journal = {J Biochem Biophys Methods},
	journal-full = {Journal of biochemical and biophysical methods},
	mesh = {4-Chloromercuribenzenesulfonate; Adolescent; Adult; Biological Transport; Body Water; Cell Membrane Permeability; Child; Child, Preschool; Erythrocyte Membrane; Erythrocytes; Female; Humans; In Vitro Techniques; Magnetic Resonance Spectroscopy; Male},
	number = {2},
	pages = {87--102},
	pmid = {2177070},
	pst = {ppublish},
	title = {On measuring the diffusional water permeability of human red blood cells and ghosts by nuclear magnetic resonance},
	volume = {21},
	year = {1990}}

@article{Bernin2013,
	doi = {10.1016/j.cocis.2013.03.007},
	author = {Bernin, Diana and Topgaard, Daniel},
	journal = {Current opinion in colloid \& interface science},
	number = {3},
	pages = {166--172},
	publisher = {Elsevier},
	title = {{NMR} diffusion and relaxation correlation methods: {N}ew insights in heterogeneous materials},
	volume = {18},
	year = {2013}}

@article{Boss2013,
	author = {Boss, Daniel and K{\"u}hn, Jonas and Jourdain, Pascal and Depeursinge, Christian and Magistretti, Pierre J and Marquet, Pierre},
	doi = {10.1117/1.JBO.18.3.036007},
	journal = {J Biomed Opt},
	journal-full = {Journal of biomedical optics},
	mesh = {Animals; CHO Cells; Cell Membrane Permeability; Cell Size; Cells, Cultured; Cricetinae; Cricetulus; Glutamic Acid; HEK293 Cells; Holography; Humans; Intracellular Space; Mice; Microscopy, Phase-Contrast; Osmosis; Osmotic Pressure; Refractometry; Reproducibility of Results; Signal Processing, Computer-Assisted; Water},
	month = mar,
	number = {3},
	pages = {036007},
	pmid = {23487181},
	pst = {ppublish},
	title = {Measurement of absolute cell volume, osmotic membrane water permeability, and refractive index of transmembrane water and solute flux by digital holographic microscopy},
	volume = {18},
	year = {2013}}

@article{Callaghan1997,
	doi = {10.1006/jmre.1997.1233},
	author = {Callaghan, P. T.},
	journal = {J. Magn. Reson.},
	pages = {74-84},
	title = {A simple matrix formalism for spin echo analysis of restricted diffusion under generalized gradient waveforms},
	volume = {129},
	year = {1997}}

@article{Callaghan2004,
	author = {Callaghan, P, T and Fur{\'o}, I},
	doi = {10.1063/1.1642604},
	journal = {J Chem Phys},
	journal-full = {The Journal of chemical physics},
	month = feb,
	number = {8},
	pages = {4032-8},
	pmid = {15268569},
	pst = {ppublish},
	title = {Diffusion-diffusion correlation and exchange as a signature for local order and dynamics},
	volume = {120},
	year = {2004}}

@article{Chakwizira2026,
	author = {Chakwizira, Arthur and {\c S}im{\c s}ek, Kadir and Szczepankiewicz, Filip and Palombo, Marco and Nilsson, Markus},
	doi = {10.1002/nbm.70315},
	journal = {NMR in Biomedicine},
	note = {e70315 NBM-25-0399},
	number = {7},
	pages = {e70315},
	title = {The Role of Dendritic Spines in Water Exchange Measurements With Diffusion {MRI}: Double Diffusion Encoding and Free-Waveform {MRI}},
	url = {https://doi.org/10.1002/nbm.70315},
	volume = {39},
	year = {2026}}

@article{Chan2025,
	author = {Chan, Kwok-Shing and Ma, Yixin and Lee, Hansol and Marques, Jos{\'e} P. and Olesen, Jonas L. and Coelho, Santiago and Novikov, Dmitry S. and Jespersen, Sune N. and Huang, Susie Y. and Lee, Hong-Hsi},
	doi = {10.1162/imag_a_00544},
	issn = {2837-6056},
	journal = {Imaging Neuroscience},
	month = {04},
	pages = {imag\_a\_00544},
	rss-description = {open access from   https://direct.mit.edu/imag/article/doi/10.1162/imag_a_00544/128696/In-vivo-human-neurite-exchange-time-imaging-at-500},
	title = {In vivo human neurite exchange time imaging at {500 mT/m} diffusion gradients},
	url = {https://doi.org/10.1162/imag_a_00544},
	volume = {3},
	year = {2025}}

@article{Dhital2019,
	author = {Dhital, Bibek and Reisert, Marco and Kellner, Elias and Kiselev, Valerij G},
	doi = {10.1016/j.neuroimage.2019.01.015},
	journal = {NeuroImage},
	journal-full = {NeuroImage},
	month = jan,
	pages = {543--550},
	pmid = {30659959},
	pst = {aheadofprint},
	title = {Intra-axonal diffusivity in brain white matter},
	volume = {189},
	year = {2019}}

@article{Fang2020,
	author = {Fang , Chengran and Nguyen, Van-Dang and Wassermann, Demian and Li, Jing-Rebecca},
	doi = {10.1016/j.neuroimage.2020.117198},
	journal = {NeuroImage},
	pages = {117198},
	title = {Diffusion {MRI} simulation of realistic neurons with {SpinDoctor} and the {Neuron Module}},
	volume = {222},
	year = {2020}}

@article{Fieremans2010,
	author = {Fieremans, Els and Novikov, Dmitry S and Jensen, Jens H and Helpern, Joseph A},
	doi = {10.1002/nbm.1577},
	journal = {NMR Biomed},
	journal-full = {NMR in biomedicine},
	mesh = {Body Water; Cell Membrane Permeability; Diffusion; Diffusion Magnetic Resonance Imaging; Humans; Models, Biological; Monte Carlo Method; Time Factors},
	month = aug,
	number = {7},
	pages = {711-24},
	pmc = {PMC2997614},
	pmid = {20882537},
	pst = {ppublish},
	title = {Monte Carlo study of a two-compartment exchange model of diffusion},
	volume = {23},
	year = {2010}}

@article{Grebenkov2007,
	doi = {10.1103/RevModPhys.79.1077},
	author = {Grebenkov, Denis S.},
	journal = {Review of Modern Physics},
	pages = {1077--1137},
	title = {{NMR} survey of reflected {B}rownian motion},
	volume = {79},
	year = {2007}}

@article{Hao2025,
	author = {Hao, Runpu and Michael, Eric Seth and Hennel, Franciszek and Pruessmann, Klaas Paul},
	doi = {10.1002/mrm.30635},
	journal = {Magnetic Resonance in Medicine},
	number = {6},
	pages = {2358-2373},
	title = {Non-monotonic time- and frequency-dependent diffusion kurtosis in the human brain revealed by pulsed and oscillating gradient experiments},
	url = {https://onlinelibrary.wiley.com/doi/abs/10.1002/mrm.30635},
	volume = {94},
	year = {2025}}

@article{Jacobs1997,
	author = {Jacobs, Bob and Driscoll, Lori and Schall, Matthew},
	doi = {10.1002/(SICI)1096-9861(19971006)386:4<661::AID-CNE11>3.0.CO;2-N},
	journal = {Journal of Comparative Neurology},
	number = {4},
	pages = {661-680},
	title = {Life-span dendritic and spine changes in areas 10 and 18 of human cortex: {A} quantitative {G}olgi study},
	url = {https://doi.org/10.1002/(SICI)1096-9861(19971006)386:4<661::AID-CNE11>3.0.CO;2-N},
	volume = {386},
	year = {1997}}

@article{Jacobs2001_neurons,
	author = {Jacobs, Bob and Schall, Matthew and Prather, Melissa and Kapler, Elisa and Driscoll, Lori and Baca, Serapio and Jacobs, Jesse and Ford, Kevin and Wainwright, Marcy and Treml, Melinda},
	doi = {10.1093/cercor/11.6.558},
	issn = {1047-3211},
	journal = {Cerebral Cortex},
	month = {06},
	number = {6},
	pages = {558-571},
	title = {Regional Dendritic and Spine Variation in Human Cerebral Cortex: a Quantitative Golgi Study},
	url = {https://doi.org/10.1093/cercor/11.6.558},
	volume = {11},
	year = {2001}}

@article{Jelescu2022_NEXI,
	author = {Jelescu, Ileana O and de Skowronski, Alexandre and Geffroy, Fran{\c c}oise and Palombo, Marco and Novikov, Dmitry S},
	doi = {10.1016/j.neuroimage.2022.119277},
	journal = {NeuroImage},
	journal-full = {NeuroImage},
	mesh = {Animals; Brain; Diffusion Magnetic Resonance Imaging; Gray Matter; Humans; Neurites; Rats; Water; White Matter},
	month = aug,
	pages = {119277},
	pmc = {PMC10363376},
	pmid = {35523369},
	pst = {ppublish},
	title = {Neurite Exchange Imaging {(NEXI): A} minimal model of diffusion in gray matter with inter-compartment water exchange},
	volume = {256},
	year = {2022}}

@article{Kaerger1985,
	author = {J{\"o}rg K{\"a}rger},
	doi = {10.1016/0001-8686(85)80018-X},
	issn = {0001-8686},
	journal = {Advances in Colloid and Interface Science},
	pages = {129-148},
	title = {{NMR} self-diffusion studies in heterogeneous systems},
	url = {https://www.sciencedirect.com/science/article/pii/000186868580018X},
	volume = {23},
	year = {1985}}

@article{Karbowski2015,
	author = {Karbowski, Jan},
	doi = {10.1371/journal.pcbi.1004532},
	journal = {PLoS Computational Biology},
	month = oct,
	number = {10},
	pages = {e1004532},
	pmcid = {PMC4593638},
	pmid = {26436731},
	publisher = {Public Library of Science},
	title = {Cortical Composition Hierarchy Driven by Spine Proportion Economical Maximization or Wire Volume Minimization},
	volume = {11},
	year = {2015}}

@article{Khateri2022,
	author = {Khateri, Mohammad and Reisert, Marco and Sierra, Alejandra and Tohka, Jussi and Kiselev, Valerij G.},
	doi = {10.1002/nbm.4804},
	journal = {NMR in Biomedicine},
	pages = {e4804},
	title = {What does {FEXI} measure?},
	year = {2022}}

@article{Lasic2011_exchange,
	author = {Lasi{\v c}, Samo and Nilsson, Markus and L{\"a}tt, Jimmy and St{\aa}hlberg, Freddy and Topgaard, Daniel},
	doi = {10.1002/mrm.22782},
	journal = {Magnetic Resonance in Medicine},
	number = {2},
	pages = {356-365},
	title = {Apparent exchange rate mapping with diffusion {MRI}},
	volume = {66},
	year = {2011}}

@article{Lee2020_GM,
	author = {Lee, Hong-Hsi and Papaioannou, Antonios and Novikov, Dmitry S and Fieremans, Els},
	doi = {10.1016/j.neuroimage.2020.117054},
	journal = {NeuroImage},
	journal-full = {NeuroImage},
	month = jun,
	pages = {117054},
	pmid = {32585341},
	pst = {aheadofprint},
	title = {In vivo observation and biophysical interpretation of time-dependent diffusion in human cortical gray matter},
	year = {2020}}

@article{Lee2025,
	author = {Lee, Hong-Hsi and Novikov, Dmitry S. and Fieremans, Els and Huang, Susie Y.},
	doi = {10.1002/mrm.30335},
	journal = {Magnetic Resonance in Medicine},
	number = {3},
	pages = {1329--1347},
	title = {Revealing membrane integrity and cell size from diffusion kurtosis time dependence},
	url = {https://onlinelibrary.wiley.com/doi/abs/10.1002/mrm.30335},
	volume = {93},
	year = {2025}}

@article{Li2019,
	author = {Jing-Rebecca Li and Van-Dang Nguyen and Try Nguyen Tran and Jan Valdman and Cong-Bang Trang and Khieu Van Nguyen and Duc Thach Son Vu and Hoang An Tran and Hoang Trong An Tran and Thi Minh Phuong Nguyen},
	doi = {10.1016/j.neuroimage.2019.116120},
	journal = {NeuroImage},
	pages = {116120},
	title = {{SpinDoctor}: {A MATLAB} toolbox for diffusion {MRI} simulation},
	volume = {202},
	year = {2019}}

@article{Li2020,
	author = {Li, Jing-Rebecca and Tran, Try Nguyen and Nguyen, Van-Dang},
	doi = {10.1002/nbm.4353},
	journal = {NMR in Biomedicine},
	note = {e4353 nbm.4353},
	number = {10},
	pages = {e4353},
	title = {Practical computation of the diffusion {MRI} signal of realistic neurons based on {L}aplace eigenfunctions},
	volume = {33},
	year = {2020}}

@article{Li2023,
	author = {Li, Chenyang and Fieremans, Els and Novikov, Dmitry S and Ge, Yulin and Zhang, Jiangyang},
	doi = {10.1002/mrm.29536},
	journal = {Magn Reson Med},
	journal-full = {Magnetic resonance in medicine},
	mesh = {Mice; Animals; Water; Magnetic Resonance Imaging; Diffusion Tensor Imaging; White Matter; Gray Matter; Diffusion Magnetic Resonance Imaging; Brain},
	month = apr,
	number = {4},
	pages = {1441--1455},
	pmc = {PMC9892228},
	pmid = {36404493},
	pst = {ppublish},
	title = {Measuring water exchange on a preclinical {MRI} system using filter exchange and diffusion time dependent kurtosis imaging},
	volume = {89},
	year = {2023}}

@article{McSweeney-Davis2025,
	author = {McSweeney-Davis, Alex and Fang, Chengran and Caruyer, Emmanuel and Kerbrat, Anne and Li, Jing-Rebecca},
	doi = {10.1093/bib/bbaf258},
	issn = {1477-4054},
	journal = {Briefings in Bioinformatics},
	month = {06},
	number = {3},
	pages = {bbaf258},
	title = {Alpha\_Mesh\_Swc: automatic and robust surface mesh generation from the skeleton description of brain cells},
	url = {https://doi.org/10.1093/bib/bbaf258},
	volume = {26},
	year = {2025}}

@article{Mitra1992,
	author = {Mitra, P.P. and Sen, P.N. and Schwartz, L.M. and Le Doussal, P.},
	doi = {10.1103/PhysRevLett.68.3555},
	journal = {Phys Rev Lett},
	journal-full = {Physical review letters},
	month = jun,
	number = {24},
	pages = {3555--3558},
	pmid = {10045734},
	pst = {ppublish},
	title = {Diffusion propagator as a probe of the structure of porous media},
	volume = {68},
	year = {1992}}

@misc{NeuroMorpho.Org,
	url = {https://neuromorpho.org},
	author = {NeuroMorpho.Org},
	howpublished = {\url{https://neuromorpho.org}},
	title = {{{NeuroMorpho.Org (RRID:SCR\_002145)}}}}

@article{Nilsson2013,
	author = {Nilsson, Markus and L{\"a}tt, Jimmy and van Westen, Danielle and Brockstedt, Sara and Lasi\v{c}, Samo and St{\aa}hlberg, Freddy and Topgaard, Daniel},
	doi = {10.1002/mrm.24395},
	journal = {Magn Reson Med},
	journal-full = {Magnetic resonance in medicine},
	mesh = {Adult; Algorithms; Brain Chemistry; Cellular Microenvironment; Diffusion; Diffusion Magnetic Resonance Imaging; Female; Humans; Image Enhancement; Image Interpretation, Computer-Assisted; Male; Reproducibility of Results; Sensitivity and Specificity; Tissue Distribution; Water},
	month = jun,
	number = {6},
	pages = {1573-81},
	pmid = {22837019},
	pst = {ppublish},
	title = {Noninvasive mapping of water diffusional exchange in the human brain using filter-exchange imaging},
	volume = {69},
	year = {2013}}

@article{Novikov2010,
	author = {Novikov, Dmitry S and Kiselev, Valerij G},
	doi = {10.1002/nbm.1584},
	journal = {NMR Biomed},
	journal-full = {NMR in biomedicine},
	month = aug,
	number = {7},
	pages = {682-97},
	pmid = {20886563},
	pst = {ppublish},
	title = {Effective medium theory of a diffusion-weighted signal},
	volume = {23},
	year = {2010}}

@article{Novikov2019_models,
	author = {Novikov, Dmitry S and Fieremans, Els and Jespersen, Sune N and Kiselev, Valerij G},
	doi = {10.1002/nbm.3998},
	journal = {NMR Biomed},
	journal-full = {NMR in biomedicine},
	mesh = {Algorithms; Anisotropy; Brain; Diffusion Magnetic Resonance Imaging; Humans; Models, Theoretical; Neurons},
	month = {04},
	number = {4},
	pages = {e3998},
	pmc = {PMC6481929},
	pmid = {30321478},
	pst = {ppublish},
	title = {Quantifying brain microstructure with diffusion {MRI}: {T}heory and parameter estimation},
	volume = {32},
	year = {2019}}

@article{Olesen2022,
	author = {Olesen, Jonas L and {\O}stergaard, Leif and Shemesh, Noam and Jespersen, Sune N},
	doi = {10.1016/j.neuroimage.2022.118976},
	journal = {NeuroImage},
	journal-full = {NeuroImage},
	mesh = {Brain; Cerebral Cortex; Diffusion Magnetic Resonance Imaging; Gray Matter; Humans; Neuroimaging; White Matter},
	month = {05},
	pages = {118976},
	pmc = {PMC8961002},
	pmid = {35168088},
	pst = {ppublish},
	title = {Diffusion time dependence, power-law scaling, and exchange in gray matter},
	volume = {251},
	year = {2022}}

@article{Ordinola2024,
	author = {Ordinola, Alfredo and {\"O}zarslan, Evren and Bai, Ruiliang and Herberthson, Magnus},
	doi = {10.1063/5.0188865},
	journal = {J Chem Phys},
	journal-full = {The Journal of chemical physics},
	month = feb,
	number = {8},
	pages = {084701},
	pmid = {38385634},
	pst = {ppublish},
	title = {Limitations and generalizations of the first order kinetics reaction expression for modeling diffusion-driven exchange: Implications on {NMR} exchange measurements},
	volume = {160},
	year = {2024}}

@article{Qiao2005,
	author = {Qiao, Ying and Galvosas, Petrik and Adalsteinsson, Thorsteinn and Sch{\"o}nhoff, Monika and Callaghan, Paul T},
	doi = {10.1063/1.1924707},
	journal = {The Journal of chemical physics},
	number = {21},
	pages = {214912},
	publisher = {American Institute of Physics},
	title = {Diffusion exchange {NMR} spectroscopic study of dextran exchange through polyelectrolyte multilayer capsules},
	volume = {122},
	year = {2005}}

@article{Quirk2003,
	author = {Quirk, James D and Bretthorst, G Larry and Duong, Timothy Q and Snyder, Avi Z and Springer, Jr, Charles S and Ackerman, Joseph J H and Neil, Jeffrey J},
	doi = {10.1002/mrm.10565},
	journal = {Magn Reson Med},
	journal-full = {Magnetic resonance in medicine},
	mesh = {Animals; Bayes Theorem; Body Water; Brain; Contrast Media; Echo-Planar Imaging; Extracellular Space; Gadolinium DTPA; Intracellular Space; Male; Rats; Rats, Sprague-Dawley; Stereotaxic Techniques},
	month = sep,
	number = {3},
	pages = {493--499},
	pmid = {12939756},
	pst = {ppublish},
	title = {Equilibrium water exchange between the intra- and extracellular spaces of mammalian brain},
	volume = {50},
	year = {2003}}

@article{Si2015,
	address = {New York, NY, USA},
	articleno = {11},
	author = {Si, Hang},
	doi = {10.1145/2629697},
	issn = {0098-3500},
	issue_date = {January 2015},
	journal = {ACM Trans. Math. Softw.},
	month = feb,
	number = {2},
	numpages = {36},
	publisher = {Association for Computing Machinery},
	title = {TetGen, a Delaunay-Based Quality Tetrahedral Mesh Generator},
	url = {https://doi.org/10.1145/2629697},
	volume = {41},
	year = {2015}}

@article{Simsek2026,
	author = {{\c S}im{\c s}ek, Kadir and Chakwizira, Arthur and Nilsson, Markus and Palombo, Marco},
	doi = {10.1002/nbm.70382},
	journal = {NMR in Biomedicine},
	note = {e70382 1674927},
	number = {10},
	pages = {e70382},
	title = {The Role of Dendritic Spines in Water Exchange Measurements With Diffusion {MRI}: Time-Dependent Single Diffusion Encoding {MRI}},
	url = {https://doi.org/10.1002/nbm.70382},
	volume = {39},
	year = {2026}}

@article{Sykova2008,
	author = {Sykov{\'a}, Eva and Nicholson, Charles},
	doi = {10.1152/physrev.00027.2007},
	journal = {Physiol Rev},
	journal-full = {Physiological reviews},
	mesh = {Animals; Brain; Brain Chemistry; Diffusion; Extracellular Space; Humans; Neuroglia; Neurons; Quaternary Ammonium Compounds},
	month = oct,
	number = {4},
	pages = {1277-340},
	pmc = {PMC2785730},
	pmid = {18923183},
	pst = {ppublish},
	title = {Diffusion in brain extracellular space},
	volume = {88},
	year = {2008}}

@article{Tecuatl2024_NeuroMorpho,
	author = {Tecuatl, Carolina and Ljungquist, Bengt and Ascoli, Giorgio A.},
	doi = {10.1096/fba.2024-00048},
	journal = {FASEB BioAdvances},
	number = {7},
	pages = {207-221},
	title = {Accelerating the continuous community sharing of digital neuromorphology data},
	url = {https://faseb.onlinelibrary.wiley.com/doi/abs/10.1096/fba.2024-00048},
	volume = {6},
	year = {2024}}

@article{Torrey1956,
	doi = {10.1103/PhysRev.104.563},
	author = {Torrey, H. C.},
	journal = {Phys. Rev.},
	pages = {563},
	title = {Bloch Equations with Diffusion Terms},
	volume = {104},
	year = {1956}}

@article{Watson2006_neurons,
	doi = {10.1016/j.neuroscience.2006.04.084},
	author = {Watson, Karli K and Jones, Todd K and Allman, John M},
	journal = {Neuroscience},
	number = {3},
	pages = {1107--1112},
	publisher = {Elsevier},
	title = {Dendritic architecture of the von {E}conomo neurons},
	volume = {141},
	year = {2006}}

@article{Williamson2019,
	author = {Williamson, Nathan H and Ravin, Rea and Benjamini, Dan and Merkle, Hellmut and Falgairolle, Melanie and O'Donovan, Michael James and Blivis, Dvir and Ide, Dave and Cai, Teddy X and Ghorashi, Nima S and Bai, Ruiliang and Basser, Peter J},
	doi = {10.7554/eLife.51101},
	journal = {Elife},
	journal-full = {eLife},
	mesh = {Action Potentials; Animals; Animals, Newborn; Anisotropy; Anterior Horn Cells; Body Water; Cell Membrane; Detergents; Deuterium; Diffusion; Diffusion Magnetic Resonance Imaging; Equipment Design; Intracellular Membranes; Magnetic Resonance Spectroscopy; Membrane Lipids; Mice; Motion; Octoxynol; Spinal Cord},
	month = dec,
	pages = {e51101},
	pmc = {PMC6977971},
	pmid = {31829935},
	pst = {epublish},
	title = {Magnetic resonance measurements of cellular and sub-cellular membrane structures in live and fixed neural tissue},
	volume = {8},
	year = {2019}}

@article{Williamson2020,
	author = {Williamson, Nathan H and Ravin, Rea and Cai, Teddy X and Benjamini, Dan and Falgairolle, Melanie and O'Donovan, Michael J and Basser, Peter J},
	doi = {10.1016/j.jmr.2020.106782},
	journal = {J Magn Reson},
	journal-full = {Journal of magnetic resonance (San Diego, Calif. : 1997)},
	mesh = {Animals; Animals, Newborn; Contrast Media; Diffusion; Equipment Design; Gadolinium DTPA; In Vitro Techniques; Magnetic Resonance Spectroscopy; Mice; Sensitivity and Specificity; Spinal Cord; Water},
	month = {08},
	pages = {106782},
	pmc = {PMC7427561},
	pmid = {32679514},
	pst = {ppublish},
	title = {Real-time measurement of diffusion exchange rate in biological tissue},
	volume = {317},
	year = {2020}}

@article{Williamson2025_preprint,
	author = {Williamson, Nathan H and Ravin, Rea and Cai, Teddy X and Rey, Julian A and Basser, Peter J},
	doi = {10.1101/2025.05.27.655493},
	elocation-id = {2025.05.27.655493},
	journal = {bioRxiv},
	publisher = {Cold Spring Harbor Laboratory},
	title = {Passive water exchange between multiple sites can explain why apparent exchange rate constants depend on ionic and osmotic conditions in gray matter},
	url = {https://www.biorxiv.org/content/early/2025/05/28/2025.05.27.655493},
	year = {2025}}

@article{Yang2018,
	author = {Yang, Donghan M and Huettner, James E and Bretthorst, G Larry and Neil, Jeffrey J and Garbow, Joel R and Ackerman, Joseph J H},
	doi = {10.1002/mrm.26781},
	journal = {Magn Reson Med},
	journal-full = {Magnetic resonance in medicine},
	month = mar,
	number = {3},
	pages = {1616--1627},
	pmc = {PMC5754269},
	pmid = {28675497},
	pst = {ppublish},
	title = {Intracellular water preexchange lifetime in neurons and astrocytes},
	volume = {79},
	year = {2018}}

@article{Zimmerman1957,
	author = {Zimmerman, J.R. and Brittin, W.E.},
	doi = {10.1021/j150556a015},
	journal = {The Journal of Physical Chemistry},
	number = {10},
	pages = {1328--1333},
	title = {Nuclear Magnetic Resonance Studies in Multiple Phase Systems: Lifetime of a Water Molecule in an Adsorbing Phase on Silica Gel},
	url = {https://doi.org/10.1021/j150556a015},
	volume = {61},
	year = {1957}}
%%%%%%%%%%%%%%%%%%

\end{document}